\documentclass[prd,a4paper,aps,showpacs,twocolumn,floats,nofootinbib,floatfix]{revtex4}

\usepackage{epsfig}
\usepackage{graphicx}
\usepackage{epstopdf}
\usepackage{color}
\usepackage{amssymb}
\usepackage{amsmath}
\usepackage{hyperref}
\usepackage{url}
\usepackage{natbib}

\newcommand{\bk}{{\mathbf k}}
\newcommand{\bq}{{\mathbf q}}
\newcommand{\bfe}{{\mathbf e}}
\newcommand{\bp}{{\mathbf p}}

\newcommand{\by}{{\mathbf y}}
\newcommand{\bx}{{\mathbf x}}

\newcommand{\bn}{{\mathbf n}}

\newcommand{\bX}{{\mathbf X}}

\newcommand{\DD}{{\cal D}}
\newcommand{\GG}{{\cal G}}

\newcommand{\HH}{{\cal H}}
\newcommand{\LL}{{\cal L}}

\newcommand{\NN}{{\cal N}}

\newcommand{\PP}{{\cal P}}
\newcommand{\SSS}{{\cal S}}
\newcommand{\XX}{{\cal X}}
\newcommand{\cd}{\cdot}

\newcommand{\eq}{{\rm{eq}}}
\newcommand{\tin}{{\rm{in}}}

\newcommand{\al}{\alpha}
\renewcommand{\b}{\beta}
\newcommand{\de}{\delta}
\newcommand{\De}{\Delta}
\newcommand{\ep}{\epsilon}
\newcommand{\ga}{\gamma}
\newcommand{\Ga}{\Gamma}

\newcommand{\La}{\Lambda}
\newcommand{\la}{\lambda}

\newcommand{\si}{\sigma}
\newcommand{\Si}{\Sigma}

\newcommand{\vph}{\varphi}
\newcommand{\ra}{\rightarrow}

\newcommand{\be}{\begin{equation}}
\newcommand{\ee}{\end{equation}}

\newcommand{\bea}{\begin{eqnarray}}
\newcommand{\eea}{\end{eqnarray}}
\newcommand{\bean}{\begin{eqnarray*}}
\newcommand{\eean}{\end{eqnarray*}}
\newcommand{\dd}{\partial}
\newcommand{\mr}{\mathrm}

\newcommand{\eg}{{\em e.g. }}

\newcommand{\lan}{\langle}
\newcommand{\ran}{\rangle}

\begin{document}

\title{Cosmic Microwave Background temperature and polarization\\ anisotropies from the large-N limit of global defects}

\author{Elisa Fenu$^{\,\dag}$, Daniel G.~Figueroa$^{\,\dag}$, Ruth Durrer$^{\,\dag}$, Juan Garc\'ia-Bellido$^{\,\S}$ and Martin Kunz$^{\,\dag,\ddag}$}

\affiliation{\vspace*{0.1cm}$^\dag$D\'epartement de Physique Th\'eorique and Center for Astroparticle Physics, 
\\ Universit\'e de 
Gen\`eve, 24 quai Ernest 
Ansermet, CH--1211 Gen\`eve 4, Switzerland}
\vspace*{0.1cm}
\affiliation{\vspace*{0.1cm}$^\S$Instituto de F\'isica Te\'orica IFT-UAM-CSIC, Universidad Aut\'onoma de Madrid,
C/ Nicol\'as Cabrera 13-15, Cantoblanco, 28049 Madrid, Spain}
\affiliation{\vspace*{0.1cm}$^\ddag$African Institute for Mathematical Sciences, 6 Melrose Road, Muizenberg,
7945, South Africa}

\date{\today}

\begin{abstract}
We determine the full $C_\ell$ spectra and correlation functions of the temperature and polarization anisotropies in the CMB, generated by a source modeled by the large $N$ limit of spontaneously broken global $O(N)$-theories. We point out a problem in the standard approach of treating the radiation-matter transition by interpolating the eigenvectors of the unequal-time correlators of the source energy-momentum tensor. This affects the CMB predictions from all type of cosmic defects. We propose a method to overcome this difficulty. We find that in the large-$N$ global model that we study, differences in the final CMB power spectra amplitudes reach $\sim 10\%-20\%$ in all channels (TT, EE, BB and TE) when compared to implementations of the eigenvector interpolation technique. We discuss as well how to optimally search for the contribution in the CMB from active sources such as cosmic defects, in experiments like Planck, COrE and PRISM.
\end{abstract}

\pacs{98.80.-k,98.80.Es}
\preprint{IFT-UAM/CSIC-13-121}

\maketitle

\section{Introduction}

The cosmic microwave background (CMB) is the most precious cosmological tool. It has not only led to two Nobel prizes in physics, but it has truly revolutionized cosmology, promoting it from an order of magnitude science to 'precision cosmology'. The reason for this is twofold: on the one hand, CMB temperature and polarization anisotropies are small, so that they can be calculated to good accuracy by linear perturbation theory and, on the other hand, very precise measurements have been performed by a range of satellite, balloon and ground based experiments~\cite{Larson:2010gs,Komatsu:2010fb,Dunkley:2010ge,Reichardt:2011yv}, most recently by the Planck collaboration~\cite{Ade:2013hta,Planck:2013kta}. The Planck temperature data demonstrates an impressive agreement with the standard flat $\Lambda$CDM model for angular scales covering three orders of magnitude, with error bars that are cosmic variance limited to above $\ell \sim 1000 $, well into the damping tail of the CMB. Planck has measured the baryon acoustic scale of the CMB to a precision of 0.06\%, and within the flat $\Lambda$CDM model it has constrained all basic parameters, with the exception of the reionization optical depth, to an accuracy of better than 3\%.

The `cleanliness' of the data and the high accuracy of the measurements render the CMB  an optimal probe of physics of the very early Universe, i.e.~at very high energy.
One suggestion, which goes back to Kibble~\cite{Kibble:1976sj,Kibble:1980mv}, is that a 
symmetry breaking phase transition in the early Universe might have led to the formation of cosmic defects. Such defects are inherently inhomogeneous and anisotropic field configurations, thus leading necessarily to fluctuations in the CMB.
{\em Local} defects are those generated from a phase transition which breaks a gauge symmetry. They only scale like the
energy density of radiation if they are line-like, i.e.~cosmic strings. Point-like local defects, e.g.~monopoles, which scale like matter and soon come to dominate the Universe, are therefore excluded. Event-like local
defects, i.e. local textures, leave no significant trace. {\em Global} defects are those from a phase transition which breaks a global symmetry. Except for the case of domain walls, which also over-close the Universe, global defects in general scale and are therefore viable, independently of their dimension. For reviews on cosmic defects see~\cite{VilenkinAndShellard,HindmarshAndKibble,Durrer:2001cg}. 

Cosmic defects lead to a variety of phenomenological effects, including the creation of CMB temperature and polarization 
anisotropies~\cite{Durrer:1994zza,Durrer2,Durrer3,Turok1,Turok2}, the imprint of non-Gaussian signatures in cosmological perturbations~\cite{Mark1,Mark2,Shellard,Dani}, the generation of cosmic rays~\cite{Hill:1986mn,KibbleCosmicRays}, or the creation of cosmic magnetic fields~\cite{Dimopoulos:1997df}. Several backgrounds of gravitational waves are also expected from the creation~\cite{DufauxFigueroaBellido}, evolution~\cite{JonesSmith:2007ne,Fenu:2009qf, Giblin:2011yh,Figueroa:2012kw} and decay~\cite{Vilenkin:1981bx,Vachaspati:1984gt, Olmez2010,Blanco-Pillado:2013qja} of cosmic defects. The amplitude of the CMB fluctuations from cosmic defects is of the order $\Delta T/T \sim 4\pi G\mu = 4\pi(M/M_p)^2$, where $M$ denotes the energy scale of the phase transition, and $G = 1/M_p^2$ is the gravitational coupling, with $M_p = 1.22\times10^{19}$ GeV the Planck mass. 

If the phase transition creating the defects is driven by thermal effects, the scale $M$ is roughly given by the critical temperature $T_c$~\cite{VilenkinAndShellard} (as long as the gauge coupling is not larger than the self-coupling of the symmetry breaking scalar field). Hence a GUT scale transition with $T_c \sim 10^{15}-10^{16}$ GeV should leave observable traces in the CMB, with $G\mu \sim 10^{-8}-10^{-6}$.  For cosmic strings, detailed simulations have led to pre-Planck constraints as of $G\mu \le 4.2\times10^{-7}$~\cite{Urrestilla:2011gr}. Assuming that cosmic string loops decay into gravitational waves,  constraints from limits on a gravitational wave background can be derived, with limits from Pulsar Timing Arrays as of $G\mu \le 5.3\times 10^{-7}$. Note however that these constraints depend on uncertain assumptions, see \cite{Binetruy:2012ze,Sanidas:2012ee} for recent discussions. Also simulations for global defects \cite{Bevis:2004wk} and for semilocal strings \cite{Urrestilla:2007sf} have been performed and have led to similar, if somewhat weaker, constraints. The current best limits from the CMB are those from the Planck collaboration \cite{Ade:2013xla}, which contend that the contribution from cosmic defects to the temperature anisotropy at multipole $l = 10$ cannot be more than $1\%-5\%$, depending on the type of defect. This translates into an improvement of the bounds to $G\mu \leq 3.0\times10^{-7}$ for Abelian-Higgs cosmic strings, to $G\mu \leq 1.3\times 10^{-7}$ for Nambu-Goto strings, and  to $G\mu \lesssim 10^{-6}$ for both semilocal strings and global $O(4)$ textures. 

It has been shown in the past~\cite{Durrer:2001xu} that the energy density from global defects is dominated by the gradient of the fields. It has also been shown that $O(N)$ models with $N>4$, which do not lead to topological defects in $3+1$ space-time dimensions, actually lead to similar results as global monopoles ($N=3$) and global textures ($N=4$). They exhibit the same scaling and the same shape of the power spectrum; when normalised to the same power at low $\ell$, their amplitudes differ by less than 30\%. The main difference is the fact that decoherence~\cite{Durrer:2001cg}, which leads to a smearing out of the acoustic peaks in the CMB power spectrum, is stronger for $N=3, 4$ defects than for the large-$N$ limit which we discuss in this paper. 

In the large-$N$ limit, $N \gg 1$, the equation of motion for the global $O(N)$ symmetric scalar field  can be linearized and solved exactly up to corrections of order $1/N$~\cite{Turok:1991qq}. This allows for an analytical understanding of the resulting non-topological field configurations. In addition, the calculation of the energy-momentum tensor and its unequal time correlators in this case only requires some  convolution integrals and no expensive numerical simulations. We will use the unequal time correlators (UTC) of the global large-$N$ limit to compute the CMB temperature and polarization anisotropies. We then compare the resulting $C_\ell$ spectra
and correlation functions with experimental capabilities and identify the best strategy to constrain defect models. Besides being a 'cheap' but quite accurate toy model for global defects, the large-$N$ limit has an interest in itself: it may very well be the case that inflation is not governed by one single scalar field but that there are multiple scalar fields which are exited e.g. during preheating. Such a situation might be modeled by the large-$N$ limit discussed in this work.

In a previous paper~\cite{GarciaBellido:2010if} we have looked at the B-polarization alone for both the large-$N$ and other defect models. In this work we discuss all the spectra and correlation functions, $TT$, $TE$, $EE$ and $BB$, but we consider only the large-$N$ model, which represents the entire class of models with several (3 or more) $O(N)$-symmetric global scalar fields. We also point out an inconsistency in the standard approach of treating the radiation-matter transition by interpolating the eigenvectors of the unequal-time correlators of the source energy-momentum tensor. We propose a method to overcome this difficulty, and characterize the differences arising in all CMB power spectra amplitudes from the large-$N$ model, as compared to previous estimations.
 
The paper is structured as follows. In section~\ref{s:spec} we discuss the large-$N$ modeling of the defects arising after the spontaneous symmetry breaking of a global O(N) theory. We place particular emphasis on the calculation of the unequal time correlators of the various energy-momentum tensor components, which are crucial for the correct computation of the the CMB signals later. In section~\ref{s:cor} we describe how to compute the CMB anisotropies and polarization amplitudes, quantifying the uncertainties 
in the calculation, which might also be relevant for cosmic strings and other defects. We calculate both the power spectra and the correlation functions. In Section~\ref{s:exp} we determine the signal to noise ratio from different observations in order to specify the optimal strategy to constrain
the model. In Section~\ref{s:con} we conclude.

{\em Notation:} Throughout we consider a flat Friedman background with metric 
$$ ds^2 =g_{\mu\nu}dx^\mu dx^\nu = a^2(t)\left[-dt^2 + \de_{ij}dx^idx^j\right]\,,$$
where $a(t)$ is the scale factor. A dot denotes a derivative w.r.t. conformal time $t$ so that $\HH = \dot a/a$ is 
the comoving Hubble parameter, related to the physical Hubble parameter $H$ by
$\HH =aH$.

\section{The large $N$ sigma-model}\label{s:spec}
\subsection{The model}

We consider an $N$-component scalar field with Lagrangian 
\be
\LL = -\dd_\mu\Phi^\dagger\dd^\mu\Phi 
-\la\left(\Phi^\dagger\Phi-v^2/2\right)^2 + \mathcal{L}_{\rm int}\ ,
\ee
where $\Phi^\dagger = (\phi_1,\phi_2,...,\phi_N)/\sqrt{2}$, and $\lambda$ and $v$ are the dimensionless self-coupling and vacuum expectation value (VEV) of $\Phi$ in the true vacuum. Here $\mathcal{L}_{\rm int}$ represents interactions of $\Phi$ with other degrees of freedom. For a phase transition within a thermal bath, $\mathcal{L}_{\rm  int}$ represents the interactions of $\Phi$ with the thermal environment at temperature $T$. In this case, and to leading order, $\mathcal{L}_{\rm int} \sim g_{\rm T}^2T^2\Phi^\dagger\Phi$, with $g_{\rm T}$ an effective thermal coupling. In the context of hybrid preheating~\cite{Felder99} $\mathcal{L}_{\rm int}$ represents interactions between $\Phi$ and a scalar singlet $\chi$, the inflaton. A typical interaction Langrangian in this scenario is $\mathcal{L}_{\rm int} = g^2\chi^2\Phi^\dagger\Phi$, where $g^2$ is a dimensionless coupling. At low temperature $T\ll v$ in one case, or small inflaton amplitude $\chi \ll (\sqrt{\lambda}/g)v$ in the other, the global $O(N)$ symmetry of the Lagrangian is spontaneously broken to $O(N-1)$. Soon after the symmetry is broken, thermal or tachyonic 
effects can be neglected, and $\Phi$ is closely confined 
(in most of space) to the vacuum manifold, given by $\Phi^\dagger\Phi = {1\over2}\sum_a\phi_a^2({\bf x},t) = {1\over2}v^2$. 
Nevertheless, in positions with co-moving distance larger than the inverse of the co-moving Hubble
parameter, $|\bx-\bx'|>\HH^{-1}$, the direction of $\Phi(\bx,t)$ and $\Phi(\bx',t)$ within the vacuum manifold are
uncorrelated due to causality. This leads to a gradient energy density associated to 
the $N-1$ Goldstone modes, $\rho \sim (\nabla\Phi)^2$. For $N > 2$, the 
dynamics of the Goldstone modes is approximately described by a non-linear sigma 
model~\cite{Turok:1991qq,Durrer:2001xu} where we enforce $\sum_a\phi_a^2 = v^2$ by a Lagrange 
multiplier. This corresponds to the limit $\la\ra\infty$ in the above Lagrangian. This approximation is very good for physical scales which are much larger than $m^{-1} \equiv 1/(\sqrt{\la}v)$. 

Normalizing the symmetry breaking field components to the VEV, $\beta_a \equiv \phi_a/v$, each component obeys the sigma model evolution equation~\cite{Kunz:1996ka}
 \begin{eqnarray}\label{e:sigma}
\Box\beta^a -(\dd_\mu\beta\cd\dd^\mu\beta)\beta^a &=&0
\end{eqnarray}
where $(\dd_\mu\beta\cd\dd^\mu\beta) =\sum_ag^{\mu\nu}\dd_\mu\beta^a(\mathbf{x},t)
\dd_\nu\beta^a(\mathbf{x},t)$ and $\sum_a\beta^a(\mathbf{x},t)
\beta^a(\mathbf{x},t) = 1$. In the large-$N$ limit, the sum 
over components can be replaced by an ensemble average over one of the field components (say the first one),
\begin{eqnarray}
\sum_a g^{\mu\nu}\partial_\mu\beta^a\partial_\nu\beta^a = N \langle g^{\mu\nu}\partial_\mu\beta^1\partial_\nu\beta^1 \rangle = \omega^2(t)\ , \label{e:ansatz}
\end{eqnarray}
where in the last equality we applied the ergodic principle, substituting ensemble averages by spatial averages. By dimensional considerations, $\omega^2(t)$ can be proportional to $\mathcal{H}^2$ and $\mathcal{H}'$ or, equivalently\,\footnote{Numerical lattice simulations of the full sigma model evolution Eq.~(\ref{e:sigma}) suggest that the ansatz (\ref{e:ansatz}) is approached on a very short timescale compared with the expansion of the Universe.},
\begin{eqnarray}
\omega^2(t) = \omega^2_ot^{-2} \,.
\end{eqnarray}
with a real and positive constant $\omega^2_o>0$. Replacing the non-linearity in the sigma-model by this expectation value
we now obtain a linear equation which can be solved exactly. In Fourier space it reads
\begin{eqnarray}\label{e:sigma2}
t^2\ddot\beta_k^{a} + 2\gamma t\dot\beta_k^{a}
  + \left(k^2t^2-{\omega_o^2}\right)\beta_k^a = 0\,,
\end{eqnarray}
where dots indicate derivatives w.r.t. conformal time $t$, and $\gamma = \frac{d\log a}{d\log t}$. In a radiation dominated 
universe $\ga=1$, while in a matter dominated universe $\ga=2$. 
The solution to Eq.~(\ref{e:sigma2}) for constant $\ga$ is given by
\begin{eqnarray}
\beta^a(k,t) = (kt)^{\frac{1}{2}-\gamma}\Big[C_1\,J_\nu(kt)+C_2\,Y_\nu(kt)\Big]\,,
\end{eqnarray}
where
\begin{eqnarray}\label{nu}
\nu^2 = \left(\frac{1}{2}-\gamma\right)^2+\omega_o^2 \ .
\end{eqnarray}
Thus, $\nu^2 > 1/4$ for a radiation dominated Universe and or $\nu^2 > 9/4$ for matter domination. 
Choosing $\nu > 0$, $Y_\nu$ diverges for small argument. We keep only the regular mode of the solution $J_\nu$, which can be written as
\begin{eqnarray}\label{e:beta}
\beta^a(k,t) &=& \sqrt{A}\left(\frac{t}{t_*}\right)^{{1\over2}-\gamma}
\frac{J_\nu(y)}{(y_*)^\nu} \beta^a(k,t_*)\,,
\end{eqnarray}
where $\beta^i(k,t_*)$ is the $i$-th component of the field at the initial 
time $t_*$ and $y=kt$, $y_*=kt_*$. In the large-$N$ limit, $\beta$ is initially distributed with 
a white noise spectrum on large scales and vanishing power on small scales
\be\label{e:coreta*}
 \langle\beta^i(\bk,t_*)\beta^j(\bk',t_*)\rangle = \left\{ 
\begin{array}{cl} 
 (2\pi)^3\frac{\de_{ij}}{N}\de(\bk+\bk') & ,\,\, kt_* \leq 1 \vspace*{0.2cm}\\
  0  & ,\,\,kt_* >1 \ . \end{array} \right.
\ee
This means that the field is aligned on scales smaller than the comoving
horizon $t_*$ and has arbitrary orientation on scales larger than $t_*$. Consistency of this solution requires~\cite{Fenu:2009qf,mkthesis}
\bea
\omega_0^2 &=& 3(\ga +1/4)\,, \qquad \nu = \ga+1  \quad \mbox{ and }\\
A &=& \frac{4\Ga(2\nu-1/2)\Ga(\nu-1/2)}{3\Ga(\nu-1)}\label{eq:normA} \ .
\eea

\subsection{Unequal time correlators}

From Eqs.~(\ref{e:beta}) and (\ref{e:coreta*}) we obtain the following 
expression for the unequal time correlators (UTCs) of the field: 
\begin{eqnarray}
&& \left\langle \beta^a(\mathbf{k},t)\beta^{*b}(\mathbf{k}',t')\right\rangle\nonumber\\
&&= A\left(\frac{tt'}{t_*^2}\right)^{3/2}\!
  \frac{J_\nu(y)J_\nu(y')}{y^\nu y'^\nu}\left\langle\beta^a(\mathbf{k},t_*)\beta^{*b}(\mathbf{k}',t_*)\right\rangle \nonumber\\
&&\equiv\! (2\pi)^3\de(\bk\!-\!\bk')\PP^{ab}_\beta(k,t,t') .
\end{eqnarray}
where 
\bea
\PP^{ab}_\beta(k,t,t') &=& \frac{\de_{ab}}{N}\frac{3A}{4\pi}(tt')^{3/2}
\frac{J_\nu(kt)J_\nu(kt')}{(kt)^\nu(kt')^\nu} \nonumber\\  &\equiv& \frac{\de_{ab}}{N}
f(k,t)f(k,t') \quad \mbox{with} \label{e:unequal}\\
f(k,t) &=& \sqrt{\frac{3A}{4\pi}}\ k^{-3/2}\frac{J_\nu(kt)}{(kt)^{\nu-3/2}}
\ .
\eea
It can be shown~\cite{Jaffe:1993tt} that in the large-$N$ limit the field $\beta$ is Gaussian distributed initially (up to corrections $\sim 1/N$). Since its evolution is linear it will remain a Gaussian field, and we can determine higher order correlators via Wick's theorem. This will be important in the next section when we determine the UTCs of its energy momentum tensor, in order to calculate the perturbations in the CMB. Notice that this source is {\em totally coherent}~\cite{Durrer:2001cg} in the sense that its UTC is a product of a function of $t$ and $t'$. Note also the $k^{-3/2}$ scaling law at horizon crossing ($kt\sim1$), analogous to the one from $de~Sitter$ quantum fluctuations. This suggests a scale-invariant spectrum of fluctuations at large scales in the CMB, just like those produced by inflation. However, since fluctuations from defects are causal they generate isocurvature, as opposed to adiabatic spectra as in inflation~\cite{Durrer:2001cg}.


In order to compute the multipolar decomposition of the CMB anisotropies and polarization variances, we need to compute the UTCs of the energy-momentum tensor of the scalar field,
\be\label{eq:Tmunu}
T_{\mu\nu}(\beta) = v^2\left[\dd_\mu\beta^a\dd_\nu\beta^a -\frac{1}{2}g_{\mu\nu}\dd_\la\beta^a\dd^\la\beta^a\right] \,.
\ee
As proposed originally in Ref.~\cite{Durrer:1994zza} and used in~\cite{Durrer:2001cg}, we parameterize $T_{\mu\nu}(\beta)$ in terms of four scalar functions,
$f_\rho$, $f_v$, $f_p$ and $f_\pi$, describing its scalar contribution to the energy density $(\rho)$, energy flux $({\rm v})$, 
pressure $(p)$ and anisotropic stress $(\pi)$ respectively; two transverse vectors, $w^{(v)}$ and $w^{(\pi)}$, describing its vector contribution
 to the energy flux and anisotropic stress and one transverse traceless tensor, $\tau^{(\pi)}$, describing the tensor anisotropic stress.
 In Fourier space these quantities are given by
 \bea
T_0^0(\beta) &=& -\frac{v^2}{a^2}f_\rho \label{eq:T00}\\
T_j^0(\beta) &=& -\frac{v^2}{a^2}\left[ik_jf_v +w^{(v)}_j\right]  \,,
   \label{eq:Tj0}\\
T_{ij}(\beta) &=& v^2\left[\de_{ij}f_p  - \left(k_ik_j-\frac{k^2}{3}\de_{ij}\right)f_\pi +\right. \nonumber \\  && \qquad \left.
\frac{i}{2}\left(w^{(\pi)}_ik_j + w^{(\pi)}_jk_i\right) +\tau_{ij}^{(\pi)}\right] \,,\label{eq:Tij}
 \eea
with
\begin{eqnarray}
k^jw^{(v)}_j &=& k^jw^{(\pi)}_j = k^j\tau_{ij}^{(\pi)} = \tau_{j}^{(\pi)\,j} = 0 \,.  \nonumber
\end{eqnarray}
Since products in real space, in Eq.~(\ref{eq:Tmunu}), turn into convolutions in Fourier space, the functions in Eqs.~(\ref{eq:T00})-(\ref{eq:Tij}) will be convolutions of powers of $\beta(\bk,t)$. Their UTCs in Fourier space can be obtained as products of the UTC of $\beta$, Eq.~(\ref{e:unequal}), and of its time derivative, using Wick's theorem. Due to the convolutions, the resulting UTCs will no longer be totally coherent. For completeness, we discuss the derivation of the energy-momentum tensor UTCs in Appendix~\ref{ap:utc}.

Using Einstein's equations, we can now determine the UTCs of the metric perturbations induced by this source. Working in longitudinal gauge, the perturbed FRW line element is given by
\bea
ds^2 &=& a^2\left[ -\left(1+2(\Psi_s+\Psi_f)\right)dt^2 + \nonumber \right.  \\ &&  \qquad \left(1-2(\Phi_s+\Phi_f)\right)\de_{ij}dx^idx^j -2\Si_idtdx^i  \nonumber \\ &&  \qquad \left.    +2h_{ij}dx^idx^j\right]\,.
\eea
Here $\Psi_s$ and $\Phi_s$ are the Bardeen potentials coming from the large-$N$ source while $\Psi_f$ and $\Phi_f$ come from the cosmic fluid (matter and radiation); $\Si_i$ and $h_{ij}$ are the vector and tensor perturbations from the the large-$N$ source, so that $\dd^i\Si_i=0$ and $h^i_i = \dd^ih_{ij} = 0$.

Setting $4\pi Gv^2/N =\ep$,  Einstein's equations give to first order in the metric perturbations,
\bea
-k^2\Phi_s &=& \ep(f_\rho + 3\HH f_v) \label{e:Phis} \\
\Psi_s-\Phi_s &=& 2\ep f_\pi   \label{e:Psis}\\
-k^2\Si_i &=& 4\ep w_i^{(v)}  \label{e:Sis} \\
\ddot h_{ij} +2\HH \dot h_{ij} +k^2h_{ij} &=& 2\ep \tau_{ij}^{(\pi)}\,.   \label{e:hs}
\eea
In a positive orthonormal frame $(\bfe^{(1)},\bfe^{(2)},\hat\bk)$, we can write $\Sigma_i$ and $h_i$ as
\bean
\Si_i = \Si_{+}\bfe^+_i +  \Si_{-}\bfe^-_i \,,~~~\\
h_{ij} = h_+\bfe^+_i\bfe^+_j  +   h_-\bfe^-_i\bfe^-_j\,, 
\eean
where $\bfe^\pm = \frac{1}{\sqrt{2}}(\bfe^{(1)}\pm i\bfe^{(2)})$, and $\Si_{\pm}$ and $h_{\pm}$ are the positive and negative helicity components of the vector and tensor contributions. For dimensional reasons and symmetry, the UTCs of these variables can be written as functions of $(y\equiv kt, y'=kt')$, or of $(z=k\sqrt{tt'},r=t'/t)$, as follows \cite{Durrer:1997ep}
\bea
\lan\varphi_i(\bk,t)\varphi^{*}_j(\bk',t')\ran = \de(\bk-\bk')\,\frac{\ep^2}{k^3z}\,R_{ij}(z,r)\,,\, \label{e:C11}  \\
\lan\Si_a(\bk,t)\Si_b^{*}(\bk',t')\ran = \de(\bk-\bk')\,\frac{\ep^2 \delta_{ab}}{k^3z}\,W(z,r)\,, \label{e:W} \\
\lan \tau^{(\pi)}_a(\bk,t) \tau^{(\pi)*}_b(\bk',t')\ran = \de(\bk-\bk')\,\frac{\delta_{ab}}{k^3z}\,T(z,r) \,, \label{e:H} 
\eea
where $a,b = \pm$, and $\vec{\varphi} = (\varphi_1,\varphi_2) \equiv (\Phi_s,\Psi_s)$. All other correlators vanish if we assume statistical homogeneity and isotropy as well as invariance under parity. The expressions of  $R_{ij},~W$ and $H$ in terms of the scalar field $\beta$ are calculated in Appendix~\ref{ap:utc}. 

The pre-factors in Eqs.~(\ref{e:C11}) to~(\ref{e:H}) have been chosen such that the remaining functions depend only on the dimensionless variables $z \equiv k\sqrt{tt'}$ and $r=t'/t$, or on $y = kt$ and $y'=kt'$. This 'scale invariance' follows from a purely dimensional argument which is strictly true only for a 'scale free' universe, e.g. during pure radiation or matter domination. As soon as a physical scale is present, as it is the case due to the transition from RD to MD at the equality time $t_{\rm eq}$, the scale invariance is broken and the correlator functions depend on $k, t$ and $t'$ separately. 

In principle, the unequal time correlators for the true expansion history of the Universe contain all the information about the large-$N$ source that we need for computing the CMB power spectra. Hence we need to compute them very carefully.

\subsection{Modeling the unequal-time correlators}
\label{sec:Method3}

Let us first consider any of our unequal time correlators from Eqs.~(\ref{e:C11})-(\ref{e:H}), which we will denote generically as $C(y,y')$. Since this is a symmetric positive operator in $y$ and $y'$ we can diagonalize it, finding an orthonormal base of eigenvectors with real positive eigenvalues $\lambda_i > 0$, which therefore can be ordered as $\lambda_1 > \lambda_2 > \lambda_3 > ... 0$. Denoting$v_n(y)$ as an eigenvector of $C(y,y')$, and $\la_n$ its
positive eigenvalue, then
\be
\int dy' g(y')C(y,y')v_n(y') =\la_nv_n(y)\,,
\ee
where $g(y')$ is a positive weight function which can be chosen appropriately. Since the eigenvectors $v_n$ are orthonormal, we have
$$ \int dy g(y) v_n(y)v_m^*(y) =\de_{nm}\,.$$
The unequal time correlator can then be written consequently as
\be
 C(y,y') =\sum_n\la_nv_n(y)v_n^*(y') \,.
\ee
In our numerical work we discretize $C(y,y')$ and order the eigenvalues such that $0<\la_{n+1}<\la_n$.

The scaling behavior, i.e.~$C(k,t,t') = C(kt,kt')$, is an extraordinarily useful property. First or all, it reduces the problem from
3 to 2 dimensions. Secondly, for $y\ll 1$ and $y'\ll 1$, $C(y,y')$ is constant. On the other hand
for $y\gg 1$ or $y'\gg 1$ it decays like a power law. This power law can be determined analytically, 
see Appendix~\ref{ap:utc}. With this we only have to determine $C(y,y')$ numerically in the regions, say
$0.1<y,\; y' < 100$.
 
In the real universe, however, we have a transition from radiation to matter domination happening shortly before decoupling. This spoils scaling. This problem arises actually for any type of cosmic defects sourcing the CMB. In the large-$N$ global scenario the index of the Bessel function in the solution for $\beta$, given by $\nu = 1+\ga$, goes from $\nu=2$ during radiation to $\nu =3$ in the matter era. However there is no analytical solution describing this transition. In the case of other defects, one often obtains the UTC's at pure RD or MD epochs alone (i.e. when there is scaling), but not in between, during the radiation-matter transition. In the literature~\cite{Turok1,Durrer:1998rw,Durrer:2001xu,Bevis:2004wk} this problem is usually dealt with by interpolating the eigenvectors from the radiation and matter dominated correlators,
\bea
\sqrt{\la_n}v_n(y,t) &=& f(t/t_{\rm eq})\sqrt{\la_n^{(r)}}v_n^{(r)}(y) + \nonumber \\
&& \left(1- f(t/t_{\rm eq})\right)\sqrt{\la_n^{(m)}}v_n^{(m)}(y)\label{eq:RDev+MDev}
\eea
where $\la_n^{(r)},~v_n^{(r)}(t)$  and  $\la_n^{(m)},~v_n^{(m)}(t)$ denote the eigenvalues and eigenvectors in the
radiation and matter dominated era respectively. Here $f(x)$ is an interpolating function verifying
\bea\label{eq:AsymtoticCond}
f(x) \xrightarrow{x\ra 0} 1~\,,~~~~~~
f(x) \xrightarrow{x\ra \infty} 0 \,.
\eea

Let us note, however,  the following problem: although the eigenvectors can be chosen real, their sign is undetermined,  they are simply rays which define a direction, but not a fixed orientation. This means that adding up linearly the eigenvector components from RD and MD sources, as in Eq.~(\ref{eq:RDev+MDev}), is not a well defined operation since the relative sign between $v_n^{(r)}$ and $v_n^{(m)}$ is undetermined. Large differences can arise in the interpolated component $v_n$ by arbitrarily flipping the sign of either $v_n^{(r)}$ or $v_n^{(m)}$.

\begin{figure}[t]
\begin{center}
\includegraphics[width=8cm]{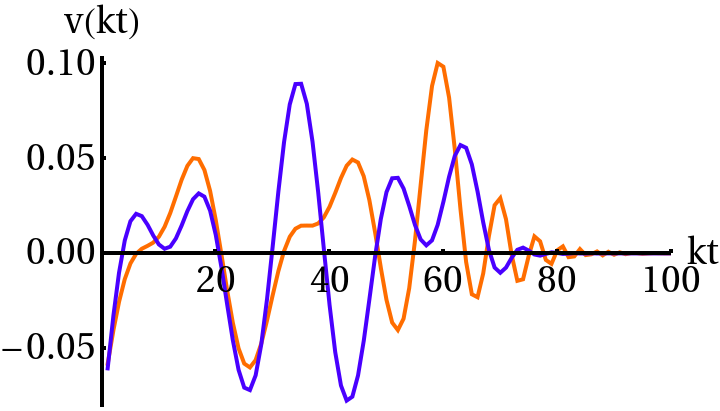}
\end{center}
\caption{14th eigenvector from of the $R_{11}$ correlator for pure RD (red) and MD (blue).}
\label{fig:RDandMD_eigenvect}
\end{figure}

One possible way to deal with this problem is to demand a positive scalar product as 
\be\label{e:pos}
\langle v_n^{(r)}v_n^{(m)}\rangle \equiv \int g(y)v_n^{(r)}(y)v_n^{(m)}(y)dy>0\,. 
\ee
After diagonalizing the RD and MD sources with arbitrary sign, one flips the sign, say, of the MD eigenvectors, in order to verify the positivity condition Eq.~(\ref{e:pos}). 

However, the eigenvectors $v^{(r)}$ and $v^{(m)}$ describe the scalar fields when these are deep in the RD and MD era, respectively. This means that some of the eigenvectors from RD and from MD will typically oscillate out of phase. In Fig.~\ref{fig:RDandMD_eigenvect} we show one eigenvector (the 14th in this example) of the unequal-time-correlator $R_{11}$, both for exact MD and RD. One can see that the two vectors are out of phase with each other for $kt \gtrsim 30$.  In this case the scale product between them becomes small, $|\langle v_n^{(r)}v_n^{(m)}\rangle| < 1$, and its sign is not very significant. 

Therefore the positive cross-product condition does not seem very meaningful, particularly because the RD and MD eigenvectors describe the sources respectively in very different epochs of the history of the Universe. 

For this reason we have considered a different approach. In particular we have introduced a procedure that does not rely on the linear superposition of the eigenvectors as in Eq.~(\ref{eq:RDev+MDev}). As we will show later, we find indeed important differences in the CMB anisotropies, of the order of few$\times 10\%$ for scalar perturbations, depending on the procedure used to determine the UTCs of the source. The issue of how to introduce correctly scaling sources in a Boltzmann code around the time of matter-radiation equality in order to obtain an accurate prediction of the CMB anisotropies, is a relevant aspect not only for the large-$N$ model, but for all scaling cosmic defects.

The origin of the problem is simple: the transition from radiation- to matter-domination breaks  scaling, i.e.~the scale free behavior of the source in the pure radiation or matter era. So the problem translates into how to source the Boltzmann equation with a scalar field  evolving around the radiation-matter equality time $t_{\rm eq}$. A  linear combination of RD and MD eigenvectors is not well defined, so we should source our code around $t_{\rm eq}$ with the physical solution for the self-ordering fields in an expanding background dictated by a mixture of radiation and matter. However, in the large-$N$ global scenario, $\beta^a(k,t)$ cannot be solved analytically in those circumstances, and secondly, it cannot be written as a function of $y = kt$. Breaking  scaling implies that the correlators depend again on the three variables, $(k,t,t')$, and not just on two $(kt,kt')$. One way to solve the problem would be to source the Boltzmann code with the UTCs calculated for each relevant $k$, as a function of $t$ and $t'$. In practice this is unfeasible.

Thus, we want to preserve the very useful property of the correlator depending only on $(kt,kt')$, while at the same time, describing correctly the evolution of the fields around $t_\eq$. Theoretically we know this is inconsistent. In practice, there is a way to circumvent the problem, as follows. Let us divide the time evolution into $q$ intervals as $t_1 < t_2 < t_3 < $ ... $ <  t_{q} < t_{q+1}$, of length $\Delta t_i = t_{i+1} - t_{i}$, $i = 1, 2, ..., q$, with $t_\eq$ lying somewhere between $t_1$ and $t_{q+1}$. If the $\Delta t_i$ intervals are sufficiently short, the behavior of the scale factor will not change appreciably between $t_i$ and $t_{i+1}$. One can then think of an adiabatic solution for the self-ordering fields within each interval $\Delta t_i$, given by Eq.~(\ref{e:beta}), but with a fixed value $\nu_i$ for the index $\nu$ between 2 and 3. Since at every time $t$ there is a well defined value of $\nu$ given by
\be\nu(t) =  1 + \left.{d\log a\over d\log t}\right|_{t}\label{eq:nuAt_t_i}\,,\ee
we can set the value of $\nu$ within the interval $(t_i,t_{i+1})$, as the arithmetic mean of the value at the boundaries,
\be \nu_{i} \equiv {1\over2}\left[\nu(t_i) + \nu(t_{i+1})\right].\label{eq:nu_i}\ee
Thus, $\nu_i$ is an effective index weighting the relative deviation from pure RD ($\nu = 2$) and MD ($\nu = 3$) during the time interval $t_i < t < t_{i+1}$, during which the adiabatic solution is written as
\be
\beta^a_{(i)}(kt) \equiv \sqrt{A_i}\left(\frac{t}{t_*}\right)^{{1\over2}-\gamma_i}
\frac{J_{\nu_i}(kt)}{(kt_*)^{\nu_i}} \beta^a(kt_*)\,,\label{eq:EffectiveBeta}
\ee
with $\gamma_i \equiv \nu_i - 1$ and $A_i$ given by the normalization constant Eq.~(\ref{eq:normA}) evaluated at $\nu = \nu_i$. By taking $q$ arbitrarily large, the set of solutions with effective indices $\nu_i$ given in Eq.~(\ref{eq:EffectiveBeta}), tend to the real physical solution. In practice we cannot take $q$ to infinity. However, if we take sufficiently small time intervals, the subsequent solutions in adjacent intervals will be similar to each other. From the computation of the UTCs with Eqs.~(\ref{e:C11})-(\ref{e:H}), in terms of convolutions of the $\beta_{(i)}^a$'s from Eq.~(\ref{eq:EffectiveBeta}), we then obtain the corresponding eigenvectors of every UTC in each interval $\Delta t_i$. The scalar product of adjacent eigenvectors will thus be large, such that the positivity condition (\ref{e:pos}) becomes meaningful again. The choice of $q$ can be made, for instance, by demanding that the total angular power spectrum $C_\ell$'s change by less than a certain tolerance, say 1\%, with increasing $q$. 

The `adiabatic' method just described should capture the evolution of the self-ordering fields with sufficient precision around $t_{\rm eq}$, ensuring an accuracy in the final $C_\ell$'s below a given tolerance requirement, while preserving at the same time the description of the UTCs as scaling functions depending on scale only through $(kt,kt')$. However theoretically correct, in practice this method is difficult to use directly. First of all, because {\em a\, priori} we do not know the number of time subintervals $q$ (for a given tolerance). This means that we must proceed by trial and error, calculating all UTCs repeatedly for every interval $t_i < t,t' < t_{i+1}$ (and from there the $C_\ell$'s), and repeating this procedure for every new set of subintervals as we increase progressively $q$. Computing all the UTCs with a good accuracy is however computationally very costly, rendering this procedure unfeasible. Secondly, the problem previously explained about the undefined sign in the method interpolating RD and MD eigenvectors as in Eq.~(\ref{eq:RDev+MDev}), is a general problem for sourcing the CMB with any type of cosmic defects. The discussed adiabatic method relies on the fact that analytical solutions exist for the self-ordering non-topological textures, but this is not the case for other defects,  particularly for the most interesting case of cosmic strings. For general defects one would need to run a large number of simulations for `intermediate' expansion rates, which again would be computationally very costly. Therefore, it would be more satisfactory to find a procedure potentially valid for any type of cosmic defects. 

Although inapplicable in practice, the previous adiabatic method still gives us the clue how to proceed. Maintaining the idea of subdividing the time evolution into $q$ intervals of length $\Delta t_i = t_{i+1} - t_{i}$, if the latter are sufficiently short, we can expect that the equal time correlators (ETCs) $C_i(k,t)$ can be written for a time $t$ within the period $(t_i,t_{i+1})$, as
\be
C_i(k,t) = f(t)C^{\rm RD}(kt) + [1-f(t)]C^{\rm MD}(kt) \,,\label{eq:C_iApprox}
\ee
where $C^{\rm RD}, C^{\rm MD}$ are the ETCs in pure RD or MD periods, and $f(t)$ is an interpolating function like in Eq.~(\ref{eq:RDev+MDev}), verifying the conditions in Eqs.~(\ref{eq:AsymtoticCond}). Since in the large-$N$ limit of global defects we can obtain $C^{\rm RD}$ and $C^{\rm MD}$ exactly from Eqs.~(\ref{e:C11})-(\ref{e:H}), just evaluating them at $t = t'$, we can then invert the problem to find $f(t)$ as
\begin{eqnarray}\label{eq:UniversalFunction}
f(t) \equiv {C_i(k,t)-C^{\rm MD}(kt)\over C^{\rm RD}(kt)-C^{\rm MD}(kt)}\,.
\end{eqnarray}
Of course, this is all under the assumption that Eq.~(\ref{eq:C_iApprox}) is a good approximation to the ETC around $t_{\rm eq}$, which is equivalent to assuming that there exists a scale-free but time-dependent interpolating function $f(t)$. In such a case, the apparent scale dependence on the $rhs.$ of Eq.~(\ref{eq:UniversalFunction}) should drop out, so that the $lhs$ is scale-independent. We should then be able to find $f(t)$ by simply computing the $rhs$ of Eq.~(\ref{eq:UniversalFunction}) for different scales $k$, and the result should always be the same at a given time $t$, independently of the scale $k$. 

\begin{figure}[t]
\begin{center}
\includegraphics[width=8cm,height=6cm,angle=0]{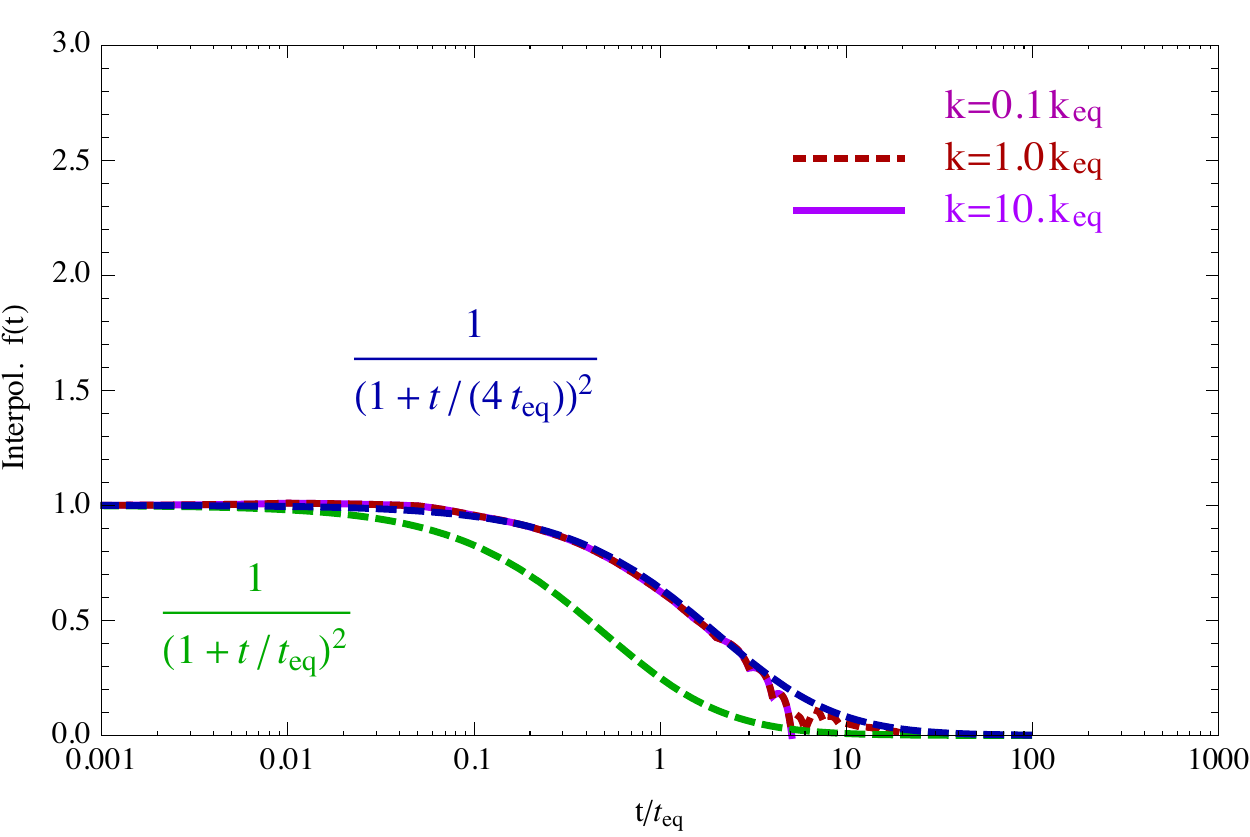}
\end{center}
\caption{The interpolation function $f(t)=(1+t/4t_{\rm eq})^{-2}$, in blue, is a universal function for all $k$. We show here the comparison for three very different wavenumbers, $k/k_{\rm eq} = 0.1,\, 1,\, 10$, around the radiation-matter transition. The alternative parametrization $f(t)=(1+t/t_{\rm eq})^{-2}$, in green, is clearly not a good description.}
\label{fig:f_t_k01_k1_k10}
\end{figure}

In order to find $f(t)$, proving at the same time its scale-invariance, we need to know the exact ETC $C_i(k,t)$ around $t_{\rm eq}$, just when the fields are not in the scaling regime and Eq.~(\ref{e:beta}) is not a valid solution. To overcome this difficulty, all we have to do is to solve numerically the equation of motion of the self-ordering fields for scales $k$ close to $k_{\rm eq}$, at times around $t_{\rm eq}$. In order to do this we need to consider the large-$N$ limit $ansatz$ Eq.~(\ref{e:ansatz}),
\begin{equation}\label{eq:omega2_ansatz}
\partial^\mu\beta_a\partial_\mu\beta_a = \omega^2(t) = c_1\mathcal{H}^2 - c_2\mathcal{H}', 
\end{equation}
but this time with the scale factor $a(t)$ given by
\begin{equation}\label{eq:ScaleFactorRDMD}
a(t) = a_{\rm eq}\left(\left[(\sqrt{2}-1)(t/t_{\rm eq})+1\right]^2-1\right)\,,
\end{equation}
which corresponds to a mixed radiation-matter fluid. We can easily fix the coefficients $c_1$ and $c_2$ by matching the expression in Eq.~(\ref{eq:omega2_ansatz}) with its asymptotic behavior in the MD and RD regimes. This yields $c_1=-3/8$ and $c_2=33/8$. Having fixed these coefficients, the mode equation can be written as
\begin{eqnarray}\label{eq:sigmaAtEq}
\ddot\beta_k^{a} + 2\mathcal{H}\dot\beta_k^{a}
  + \left(k^2 - \omega^2(t)\right)\beta_k^a = 0\,,
\end{eqnarray}
with $\omega^2(t)$ and $a(t)$ given by Eq.~(\ref{eq:omega2_ansatz}) and Eq.~\ref{eq:ScaleFactorRDMD}, respectively. We have solved Eq.~(\ref{eq:sigmaAtEq}) for $k/k_{\rm eq} = 0.1, 1$ and $10$, for a large time interval $t = 10^{-3}t_{\rm eq}-10^{3}t_{\rm eq}$. From there we have computed the ETCs evaluating Eqs.~(\ref{e:C11}) to~(\ref{e:H}) with the numerical solutions, and obtained the function $f(t)$ for each scale $k$ considered, via Eq.~(\ref{eq:UniversalFunction}). The result is shown in Fig.~\ref{fig:f_t_k01_k1_k10}. Clearly the interpolating function, $f(t)$, is the same for every scale $k$. We fitted the curves with a function $f$ given by
\begin{equation}\label{eq:InterpolationFunct}
f(t) = \left[1 + (t/4t_{\rm eq})\right]^{-2}\,,
\end{equation} 
which does an excellent job. 

If the subintervals are short enough (say $\Delta t_i \ll t_i, t_{i+1}$ and therefore $t,t' \sim t_{i},t_{i+1}$), then we should also be able to approximate each UTC at the times $t_i < t,t' < t_{i+1}$, by\footnote{An alternative approach would have been to replace $f(\bar t_i)$ by $\sqrt{f(t_{i})f(t_{i+1})}$, but in principle there is no more reason for one choice or another.}
\bea
C_i(k,t,t') &=& f(\bar t_i)C^{\rm RD}(kt,kt')\nonumber\\
 && + \left[1-f(\bar t_i)\right]C^{\rm MD}(kt,kt')\,,\label{eq:C_iApproxUTC}
\eea
with $\bar t_{i} \equiv (t_i + t_{i+1})/2$, and $C^{\rm X}$ the UTCs calculated with solution Eq.~(\ref{e:beta}) for $\nu = 2$ (X = RD) or $\nu = 3$ (X = MD). 
The larger is $q$ the shorter are the time intervals, and therefore the more accurate this ansatz approaches the real physical answer. To test the approximation, we simply require the same criteria stated before for the adiabatic approximation: the total $C_\ell$'s obtained from a given $q$ should change by less than a given percent tolerance when we increase the number of subintervals. We hope that our approximation reproduces the physical solution in that moment, with an accuracy better than the chosen tolerance.

After trial and error, we have found that we satisfy the above criterion for a $1\%$ tolerance, by taking $q = 11$ and choosing the boundary times $t_i$ in the intervals as follows: Evaluating Eq.~(\ref{eq:nuAt_t_i}) with the scale factor Eq.~(\ref{eq:ScaleFactorRDMD}), we have considered regularly spaced values (except for the extreme values $t_1$ and $t_q$) of the  effective index $\nu$ of the adiabatic approximation, $\nu(t_2) = 2.05, \nu(t_3)= 2.15, \nu(t_4) = 2.25, ..., \nu(t_{10}) = 2.85, \nu(t_{11}) = 2.95$, and $\nu(t_1) = 2.01$ and $\nu(t_{12}) = 2.99$. From here we find the times at the boundaries of the intervals by inverting the relation~(\ref{eq:nuAt_t_i}), yielding
$$
\begin{array}{cc}
\nu(t_{\,1}) = 2.01\,, & t_{\,1}/t_\eq = 0.049\\
\nu(t_{\,2}) = 2.05\,, & t_{\,2}/t_\eq = 0.254\\
\nu(t_{\,3}) = 2.15\,, & t_{\,3}/t_\eq = 0.852\\
\nu(t_{\,4}) = 2.25\,, & t_{\,4}/t_\eq = 1.609\\
\nu(t_{\,5}) = 2.35\,, & t_{\,5}/t_\eq = 2.600\\
\nu(t_{\,6}) = 2.45\,, & t_{\,6}/t_\eq = 3.950\\
\nu(t_{\,7}) = 2.55\,, & t_{\,7}/t_\eq = 5.901\\
\nu(t_{\,8}) = 2.65\,, & t_{\,8}/t_\eq = 8.967\\ 
\nu(t_{\,9}) = 2.75\,, & t_{\,9}/t_\eq = 14.49\\
\nu(t_{10}) = 2.85\,, & t_{10}/t_\eq = 27.36\\
\nu(t_{11}) = 2.95\,, & t_{11}/t_\eq = 91.74\\
\nu(t_{12}) = 2.99\,, & t_{12}/t_\eq = 478.0
\end{array}
$$

For $t < t_1$, we provide just the UTCs  from the  pure RD universe, $\nu=2$, whereas for $t > t_{12}$, we use the MD solution, $\nu=3$. Note however that the present age of the Universe $t_0$ is actually smaller than $t_{12}$, so in practice, we never source the Boltzman code with the solution from a pure MD universe. Note also that we do not discuss the fact that the Universe becomes $\La$--dominated at late times, since there is no need: in that case the linearized field equation for $\beta^a$ is not of the form~(\ref{e:sigma}). The term $\ga t$ has to be replaced by $\HH t^2$, which spoils scaling and the possibility of obtaining an analytic solution. However it is expected that the main effect will appear at the quadrupole and octopole moments of the power spectrum, whose uncertainty is dominated by cosmic variance.

As mentioned before, we have found that $q = 11$ is the minimum number of subintervals required for the total $C_\ell$'s to change by less than $1\%$ when subdividing the time evolution with one more subinterval, $q \rightarrow q+1$. There is the possibility, however, that when increasing further the number of subintervals, the accumulated change could become larger than $1\%$ with respect to the case $q = 11$. In order to avoid this, we have yet imposed a stronger criteria: that the $C_\ell$'s should not change more than $1\%$ when increasing the number of subintervals as $q \rightarrow 2q$. We have considered the following times  
$$
\begin{array}{cc}
\nu(t_{\,1}) = 2.01\,, & t_{\,1}/t_\eq = 0.049\\
\nu(t_{\,2}) = 2.03\,, & t_{\,2}/t_\eq = 0.150\\ 
\nu(t_{\,3}) = 2.06\,, & t_{\,3}/t_\eq = 0.308\\
\nu(t_{\,4}) = 2.10\,, & t_{\,4}/t_\eq = 0.536\\
\nu(t_{\,5}) = 2.15\,, & t_{\,5}/t_\eq = 0.852\\
\nu(t_{\,6}) = 2.20\,, & t_{\,6}/t_\eq = 1.207\\
\nu(t_{\,7}) = 2.25\,, & t_{\,7}/t_\eq = 1.609\\
\nu(t_{\,8}) = 2.30\,, & t_{\,8}/t_\eq = 2.069\\ 
\nu(t_{\,9}) = 2.35\,, & t_{\,9}/t_\eq = 2.600\\
\nu(t_{10}) = 2.40\,, & t_{10}/t_\eq = 3.219\\ 
\nu(t_{11}) = 2.45\,, & t_{11}/t_\eq = 3.950\\
\nu(t_{12}) = 2.50\,, & t_{12}/t_\eq = 4.828\\
\nu(t_{13}) = 2.55\,, & t_{13}/t_\eq = 5.901\\
\nu(t_{14}) = 2.60\,, & t_{14}/t_\eq = 7.243\\
\nu(t_{15}) = 2.65\,, & t_{15}/t_\eq = 8.967\\ 
\nu(t_{16}) = 2.70\,, & t_{16}/t_\eq = 11.27\\ 
\nu(t_{17}) = 2.75\,, & t_{17}/t_\eq = 14.49\\
\nu(t_{18}) = 2.80\,, & t_{18}/t_\eq = 19.31\\
\nu(t_{19}) = 2.85\,, & t_{19}/t_\eq = 27.36\\
\nu(t_{20}) = 2.90\,, & t_{20}/t_\eq = 43.46\\
\nu(t_{21}) = 2.95\,, & t_{21}/t_\eq = 91.74\\
\nu(t_{22}) = 2.97\,, & t_{22}/t_\eq = 156.1\\
\nu(t_{23}) = 2.99\,, & ~~t_{23}/t_\eq = 478.0\,,
\end{array}
$$
as the boundaries of $q = 22$ subinterval around $t_{\rm eq}$, again regularly spaced in $\nu$ (except for close to the extremes). We have found that indeed when increasing the number of subdivisions to $q = 22$, the total $C_\ell$'s do not change by more than $1\%$ with respect the corresponding amplitudes obtained for $q = 11$. 

\begin{figure}[t]
\begin{center}
\includegraphics[width=8cm]{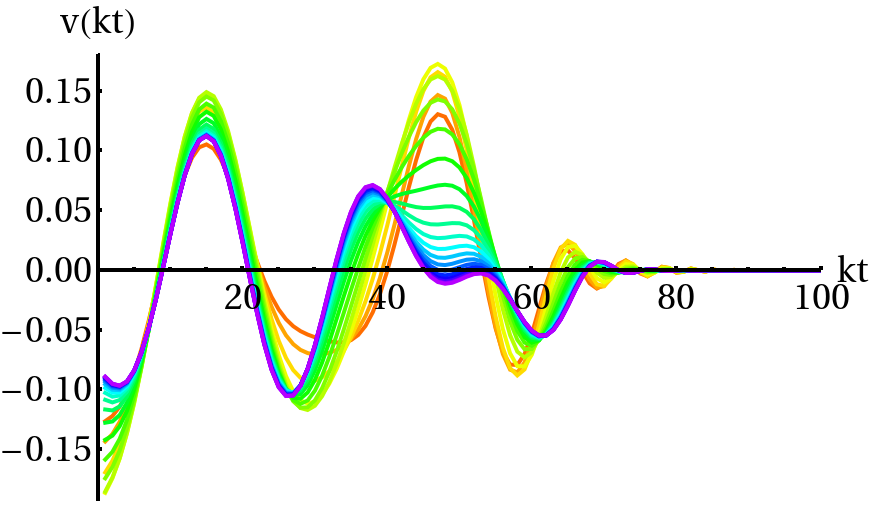}
\end{center}
\caption{The 11th eigenvector from the $C_{i}$ tensor correlators within the succesive periods $(t_i,t_{i+1})$. The color coding shows the transition from red ($\nu = 2.03$) to orange ($\nu = 2.10$), yellow and green(s) ($\nu = 2.15-2.50$), blue(s) ($\nu = 2.55 - 2.95$) and finally purple ($\nu = 2.97$). }
\label{fig:RDtoMD_eigenvectEvolution}
\end{figure}

In what follows we show the results from $q = 22$ subintervals with the time boundaries listed above, since these are the most precise calculations we have done. We however insist on the fact that these spectra  differ by less than $1\%$ from the ones obtained with $q = 11$, and that showing the latter would had sufficed as well. We have computed the matrices $C^{\rm RD}(y,y')$ and $C^{\rm MD}(y,y')$ for all sources (scalar, vector and tensor) with a high resolution integrator (4 months computation in a standard CPU serial processor). Then we have built the corresponding correlators $C_i(y,y')$ for every interval $t_i < t < t_{i+1}$, by means of Eq.~(\ref{eq:C_iApproxUTC}) with $f(t) = (1+t/4t_\eq)^{-2}$ evaluated at the intermediate times $\bar{t}_{i} \equiv (t_i + t_{i+1})/2$.

We have diagonalized the scalar, vector and tensor correlators $C_i$'s (defining $C_0$ as $C_{\rm RD}$) and sourced the Boltzman code at the times $t_i < t < t_{i+1}$ with the corresponding eigenvectors $v_n^{(i)}$. In order to match smoothly the eigenvectors from a  correlator $C_{i-1}$ with those from $C_{i}$ at the transition times $t = t_{i}$, we have imposed the positivity criterion,
$$\delta_{n,i} \equiv \int g(y)v_n^{(i-1)}(y)v_n^{(i)}(y)dy > 0 \,.$$ 
This criterion becomes now always meaningful, since the time subintervals are sufficiently short so that the $n$th eigenvector of $C_{i}$ is only 'slightly' out of phase with respect to the corresponding one of $C_{i-1}$. This is opposite to  matching eigenvectors from deep in RD with those deep in MD, which are significantly out of phase, see Fig.~\ref{fig:RDandMD_eigenvect}. In Fig.~\ref{fig:RDtoMD_eigenvectEvolution} we see how an eigenvector $v_n^{(i)}$ changes smoothly to $v_n^{(i+1)}$. In particular, we are plotting the 11th eigenvector from the tensor UTCs $H^{(i)}$ obtained at each period $(t_i,t_{i+1})$. The color coding shows the transition from red, corresponding to the closest one to pure RD with an effective index $\nu = 2.01$, to orange ($\nu = 2.10$), then yellow and different greens for $\nu = 2.15-2.50$, different blues for $\nu = 2.55-2.95$, and finally purple, corresponding to the one closest to MD, $\nu = 2.97$. 

Note also that for cosmic string simulations one usually computes their UTC's in the scaling regimes in pure RD and MD, and then interpolate the corresponding eigenvectors as in Eq.~(\ref{eq:RDev+MDev})~\cite{Bevis:2004wk}. Our exercise shows that one gets significantly different results in the CMB power spectra, see next section, when one compares the method we have proposed versus the standard interpolation method at the level of the eigenvectors. It would be therefore very interesting to repeat this exercise with cosmic string UTCs. It is possible that the interpolation function that we have found, Eq.~(\ref{eq:InterpolationFunct}), is universal, in the sense that it can be used for any type of defect. However, we have found its time dependence from the large-$N$ model, by solving numerically the scalar field evolution around $t_{\rm eq}$. Thus, although considering it as a plausible speculation that $f(t)$ given by Eq.~(\ref{eq:InterpolationFunct}) might be the one to be used for every type of defects  -- why should it depend on the large-$N$ model? -- this can only be demonstrated with defect simulations around $t_{\rm eq}$, which is beyond the scope of this paper. Despite the absence of this exercise, we suggest the use of our $f(t)$ for other defects as well.
Besides, the new method described by Eq.~(\ref{eq:C_iApproxUTC}) should of course replace the old eigenvector interpolation prescription for introducing active sources in CMB codes. 

\section{CMB power spectra and correlation functions}\label{s:cor}
\subsection{The formalism\label{s:formalism}}
Formally, the CMB spectra are of the form
\be
C_\ell^{XY} = \int \frac{dk}{k} \De_\ell^{XY}(k) \,, 
\ee
where $X$ and $Y$ are $T$, $E$ or $B$, and
\begin{eqnarray}
\lan X_\ell(\bk,t_0)Y_{\ell}(\bk,t_0)\ran = (2\pi)^3\de(\bk-\bk')\De_\ell^{XY} .
\end{eqnarray}
The only non-vanishing cross correlation is $TE$, both $TB$ and $EB$ vanish in a universe which is invariant under parity.  
Collecting all the perturbation variables $X_\ell$ as well as the dark matter, the baryon and the neutrino perturbations into one long vector which we call $\bX(\bk,t)$, the first order perturbation equation is of the form
\be\label{e:dif1}
\DD_{ij} X_j(\bk,t)  =\SSS_i(\bk,t) \,.
\ee
Here $\DD_{ij}$ is a first order differential operator depending on time and $\SSS_i$ is the source which can be parameterized 
in terms of $\Phi_s,$ $\Psi_s$, $\Si_i$ and $h_{ij}$. Be $\GG_{ij}(k,t,t')$ the Green function for $\DD_{ij}$ which depends only on the background universe. Then the solution with vanishing initial condition at $t_*$ is given by
\be\label{e:solgen}
X_i(\bk,t) = \int_{t_*}^tdt'\GG_{ij}(k,t,t')\SSS_j(\bk,t') \,,
\ee
and the two point correlators are
\bea
&&\lan X_i(\bk,t)X_j^*(\bk',t)\ran =    \\
 &&\hspace*{-0.9cm} \int_{t_*}^t\!dt'dt''\GG_{im}(k,t,t')\GG_{jn}(k',t,t'')\lan\SSS_m(\bk,t')\SSS_n(\bk',t'')\ran\nonumber 
\eea

If we diagonalize the UTCs of the source as
\bea
&& \lan\SSS_m(\bk,t')\SSS_n(\bk',t'')\ran \\ \nonumber
&& = (2\pi)^3\de(\bk-\bk')\sum_p\la_pv_m^{(p)}(k,t')v_n^{(p)}(k,t'')\,,
\eea
we obtain
\bea
&&\lan X_i(\bk,t)X_j^*(\bk',t)\ran =    (2\pi)^3\de(\bk-\bk')\sum_p\la_p\times  \nonumber \\
 &&\hspace*{-0.9cm} \int_{t_*}^t\!dt'dt''\GG_{im}(k,t,t')\GG_{jn}(k,t,t'')v_m^{(p)}(k,t')v_n^{(p)}(k,t'')\,. 
\eea
The power spectra evaluated today, which are defined by 
$$\lan X_i(\bk,t_0)X_j^*(\bk',t_0)\ran =   (2\pi)^3\de(\bk-\bk')P_{ij}(k)$$
are then given as by a sum of products of deterministic (not stochastic) solutions,
\bea
&& \hspace{-0.7cm} P_{ij}(k) =\sum_mP_{ij}^{(m)}(k) = \sum_m\XX_i^{(m)}\XX_j^{*\,(m)}(k)  \mbox{ with}\\
&& \XX_i^{(m)}(k) = 
 \sqrt{\la_m}\int_{t_*}^{t_0}\!dt\GG_{ij}(k,t_0,t)v_j^{(m)}(k,t)\,. 
 \eea
Hence $P_{ij}^{(m)}(k)$ is the product of the solutions of Eq.~(\ref{e:dif1}) with source $\sqrt{\la_m}v_i^{(m)}$
and  $\sqrt{\la_m}v_j^{(m)}$ respectively.
This explains, why the unequal time correlators are all we need to calculate the power spectra within linear perturbation theory. For more details, see Ref.~\cite{Durrer:2001cg}.

\begin{figure}[t]
\begin{center}
\includegraphics[width=8cm,height=6cm,angle=0]{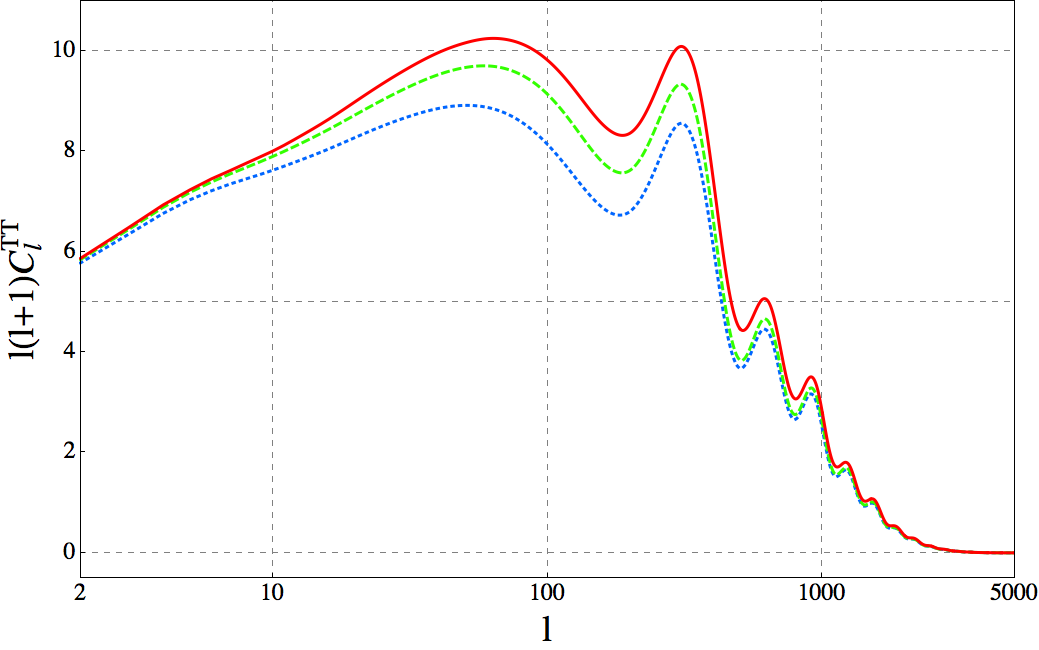}
\end{center}
\caption{The total $C_{\ell}^{TT}$ spectrum (sum of scalar, vector and tensor contributions) from procedure 1 (dotted, blue), 2 (dashed, green) and 3 (solid, red). Note the different position of the acoustic peaks as compared to the standard inflationary spectrum, e.g.~the first peak is at $\ell \sim 50$ versus the usual $\ell \sim 200$.
}
\label{fig:totalTT}
\end{figure}

\begin{figure}[t]
\begin{center}
\includegraphics[width=8cm,height=6cm,angle=0]{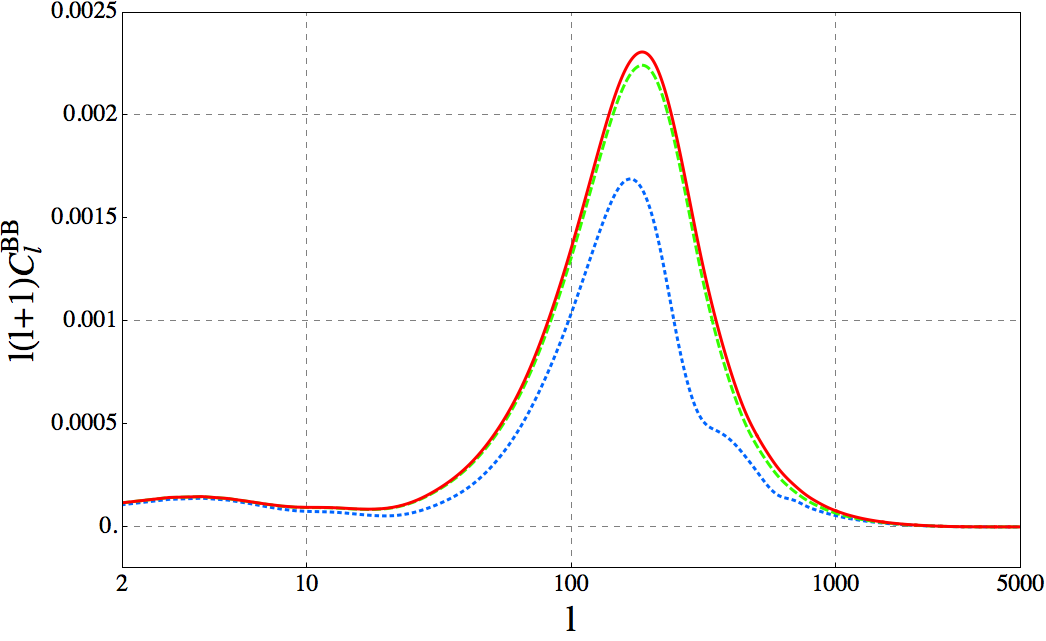}
\end{center}
\caption{The total $C_{\ell}^{BB}$ spectrum (sum of scalar, vector and tensor contributions) from procedure 1 (dotted, blue), 2 (dashed, green) and 3 (solid, red). 
}
\label{fig:totalBB}
\end{figure}

\begin{figure}[t]
\begin{center}
\includegraphics[width=8cm,height=6cm,angle=0]{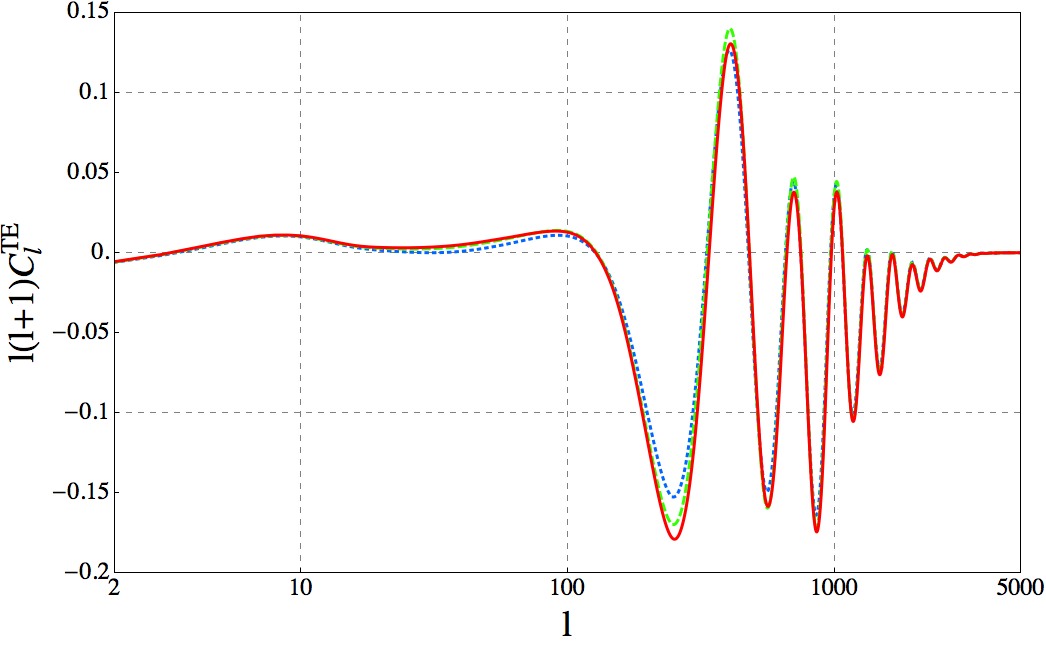}
\end{center}
\caption{The total $C_{\ell}^{TE}$ spectrum (sum of scalar, vector and tensor contributions) from procedure 1 (dotted, blue), 2 (dashed, green) and 3 (solid, red). 
}
\label{fig:totalTE}
\end{figure}

\begin{figure}[t]
\begin{center}
\includegraphics[width=8cm,height=6cm,angle=0]{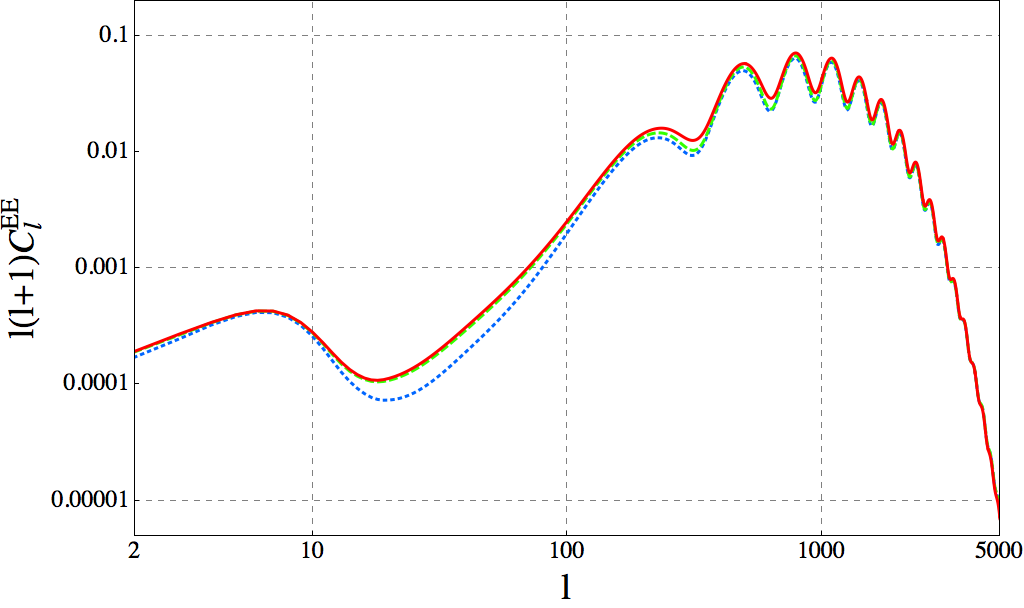}
\end{center}
\caption{The total $C_{\ell}^{EE}$ spectrum (sum of scalar, vector and tensor contributions) from procedure 1 (dotted, blue), 2 (dashed, green) and 3 (solid, red). 
}
\label{fig:totalEE}
\end{figure}

\subsection{The power spectra}

We have used a modified version of CMBEASY~\cite{Doran:2003sy} to include sources~\cite{Bevis:2004wk}. We have then computed the CMB power spectra from large-$N$ global defects using various procedures. First of all, we have obtained the CMB spectra using Eq.~(\ref{eq:RDev+MDev}) in two different ways, which we refer to as procedures 1 and 2: In procedure 1, we use the eigenvectors with arbitrary sign as given by default by the diagonalization algorithm of the correlators. In procedure 2, we use the same eigenvectors but only after having flipped the signs appropriately, such that the positivity criterion~(\ref{e:pos}) is imposed between MD and RD eigenvectors. On the other hand, we have also obtained the CMB power spectra by using the procedure explained in detail in section~\ref{sec:Method3}, which we will refer to as procedure 3 from now on. By this we mean that we have divided the time evolution into $q$ sub-intervals $(t_i,t_{i+1})$, $i = 1, 2,$ ... $q$, and then we have introduced as a source at each interval the eigenvectors of the UTCs given by Eq.~(\ref{eq:C_iApproxUTC}). We consider this latter procedure as the closest one to the physical answer. We have varied the number of intervals until a further increase changes the resulting CMB spectra by less than a given tolerance factor, which we fixed as 1\%.

In the first series of plots, Figs.~\ref{fig:totalTT}-\ref{fig:totalEE}, we compare the shape and amplitude of the different CMB power spectra obtained by the three different procedures. The color-coding/line-style among them is shared, with blue/dotted for procedure 1, green/dashed for procedure 2, and red/solid for procedure 3. In these figures we show the total amplitude for the TT, BB, EE and TE channels, respectively, having summed up in each channel the corresponding contribution from the first 200 eigenvectors of all perturbations (scalar, vector and tensor). In the TT anisotropies, see Fig.~\ref{fig:totalTT}, the difference in amplitude between the three methods reaches up to about $25\%$ in the height of the first peak when comparing procedure 1 with procedure 3. The amplitude of the spectrum obtained with procedure 1 is of course random to some extent, since the relative sign between MD and RD eigenvectors used is random. But even when comparing the output from procedure 2 with that from procedure 3, the difference in amplitude is still of the order of $10\%-15\%$. For the BB channel, the differences between procedures 1 and 3 reach $\sim 100\%$ (i.e.~a factor $\sim 2$ of discrepancy), but on the other hand, the difference when comparing procedures 2 and 3 only amounts to a 2-3 \% at low $\ell$ (although it goes up to $10\%-15\%$ for $\ell > 300$, a feature not appreciated by eye in the linear plot in Fig.~\ref{fig:totalBB}). In the TE channel the difference in amplitude between procedure 3 with respect both procedures 1 and 2, reaches about $10\%-15\%$ (when comparing the curves far from the zeros of $C_\ell^{\rm TE}$). In the case of EE, the relative amplitude between procedures 1 and 3 is of the order of a few times $10\%$ (reaching even $\sim 70\%$ at $\ell \simeq 20-30$), while it becomes smaller when comparing the amplitudes from procedures 2 and 3, differing $2\%-6\%$ for $\ell < 100$, but reaching up to $\sim 20\%$ in the dips of the oscillations at multipoles $\ell \ge 200$.

\begin{figure}[t]
\begin{center}
\includegraphics[width=8cm,height=6cm,angle=0]{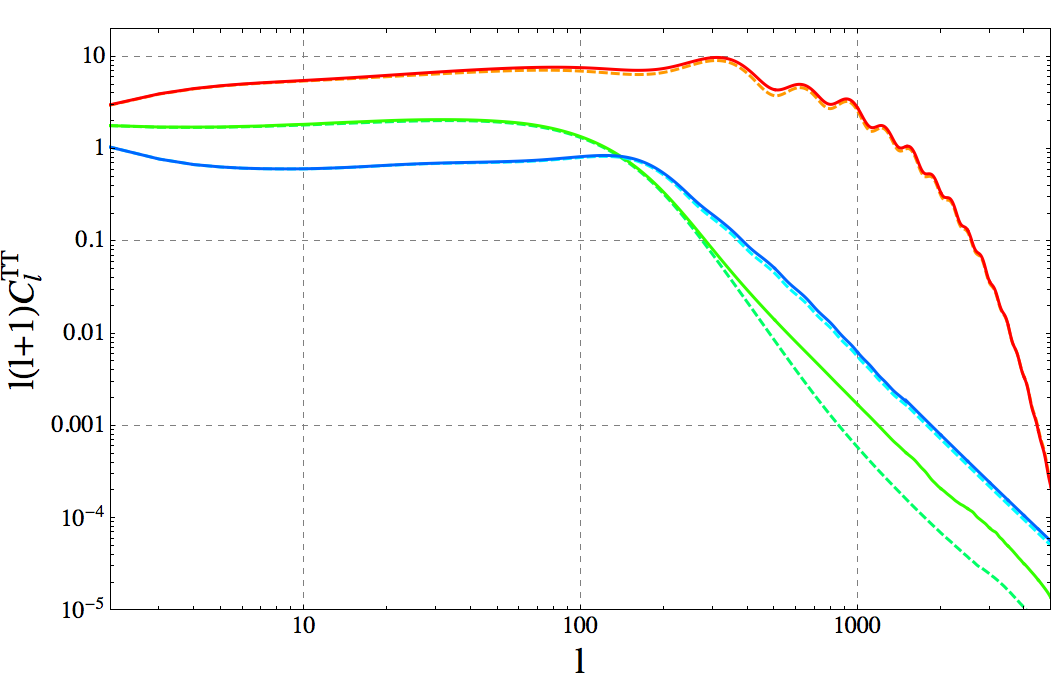}
\end{center}
\caption{The $C_\ell^{TT}$ spectrum decomposed into its scalar (red/orange top curves at low $\ell$), vector (green middle curves at low $\ell$) and tensor (blue bottom curves at low $\ell$) parts. Here, and in the analogous figures for $BB$, $TE$ and $EE$, we only show the differences between procedures 2 (dashed lines) and 3 (solid lines). In this case the differences are most significant for the dominant scalar perturbations, whilst they are small for vector perturbations (except at very small scales $\ell > 400$, where however vectors are completely subdominant).}
\label{fig:partialTT}
\end{figure}

Clearly there are noticeable differences in amplitude depending on the procedure used to treat the defect sources. Using a linear combination of RD and MD eigenvectors  is not well defined (procedures 1 and 2) due to the sign choice, which also in procedure 2 is still somewhat arbitrary. Our procedure 3 is more realistic. With the choice of time subintervals discussed in the previous section, an accuracy of order $\sim 1\%$ is reached in the final answer. As shown in  Figs.~\ref{fig:totalTT}-\ref{fig:totalEE}, differences of order $\mathcal{O}(10) \%$ arise in the channels TT, TE and EE, and of order $\mathcal{O}(1) \%$ in the channel BB, when comparing the amplitudes obtained with (the more physically correct) procedure 3 versus the procedure 2. This difference is not relevant from the point of view of constraining the symmetry scale. The UTCs, and therefore the $C_\ell$'s, scale as (VEV)$^4$ and therefore the differences found in the temperature and polarization power spectra will translate at most into a few \% difference in the upper bound for the VEV, which does not represent a significant improvement. However, from the point of view of detecting defects in the CMB, the differences found are relevant, since they depend on the multipole $\ell$ and therefore they also change the resulting shapes of the spectra. For instance, in Fig.~\ref{fig:totalEE} one can observe how the relative amplitude of the first valley at $\ell \approx 300$ with respect to the amplitude of the valley at $\ell \approx 20$, is higher than in the procedures 1 and 2. In other words, using procedures 1 or 2 we would be looking for a signal with the second trough ($\ell \approx 300$) at a given relative amplitude with respect to the first one ($\ell \approx 20$), but we find with procedure 3 that the effect of considering the field evolution around $t_\eq$ in a more precise manner, lifts up the second trough with respect to the first one.

\begin{figure}[t]
\begin{center}
\includegraphics[width=8cm,height=6cm,angle=0]{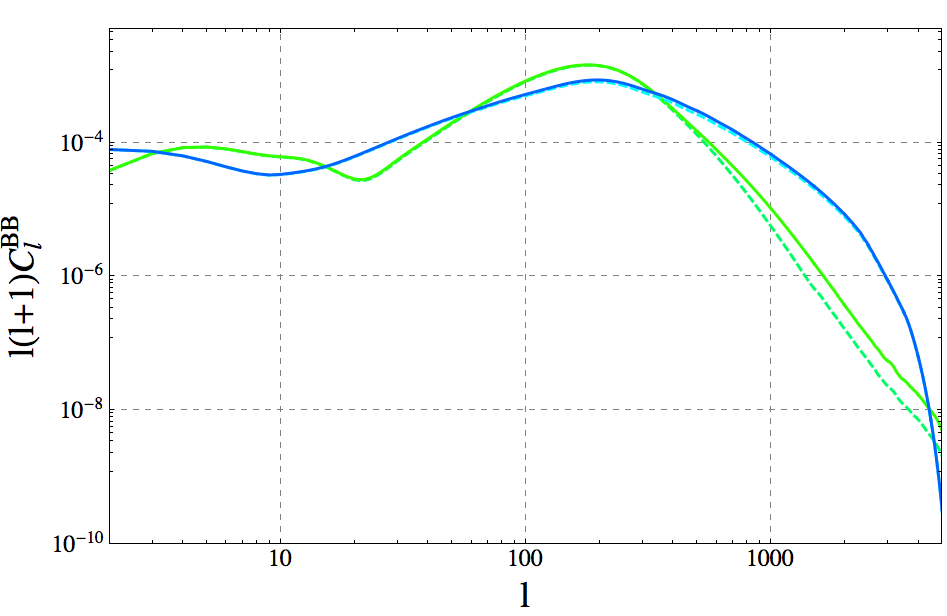}
\end{center}
\caption{The $C_\ell^{BB}$ spectrum decomposed into its vector (green (upper) curves at $\ell = 100$) and tensor (blue (lower) curves at $\ell = 100$) parts. In this channel there are no scalar perturbations, and the differences between procedures 2 and 3 are of the same order for the tensor and vector contributions.}
\label{fig:partialBB}
\end{figure}

\begin{figure}[t]
\begin{center}
\includegraphics[width=8cm,height=6cm,angle=0]{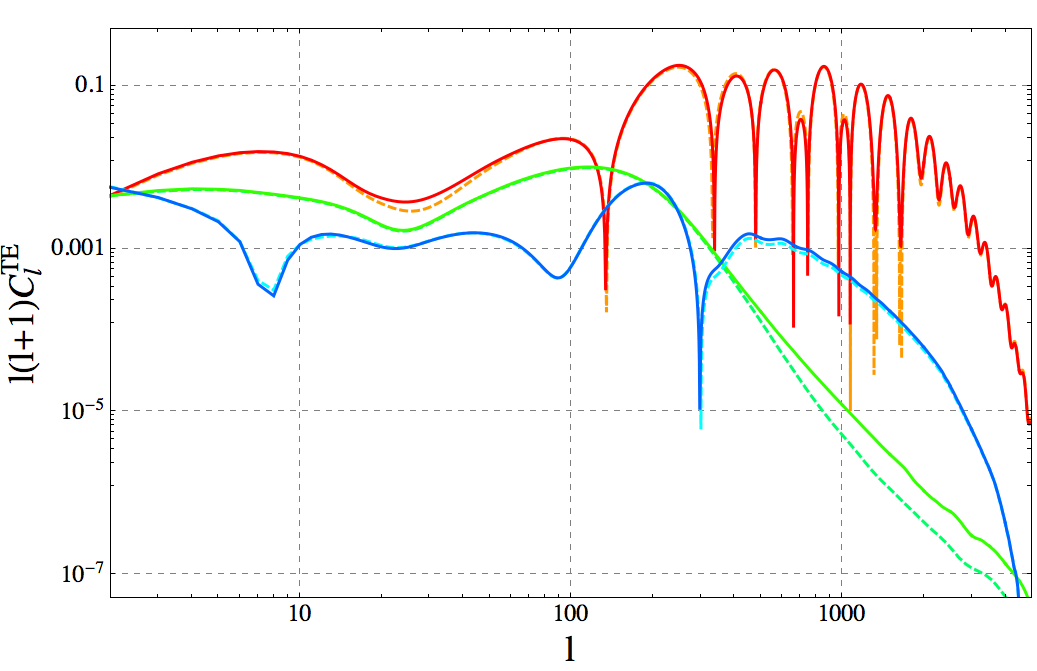}
\end{center}
\caption{The $C_\ell^{TE}$ spectrum decomposed into its scalar (top red/orange curves at $\ell = 10$), vector (middle green curves at $\ell = 10$) and tensor (bottom blue curves at $\ell = 10$) parts. 
 Again, the differences between procedures 2 and 3 are small for vector perturbations.}
\label{fig:partialTE}
\end{figure}

\begin{figure}[t]
\begin{center}
\includegraphics[width=8cm,height=6cm,angle=0]{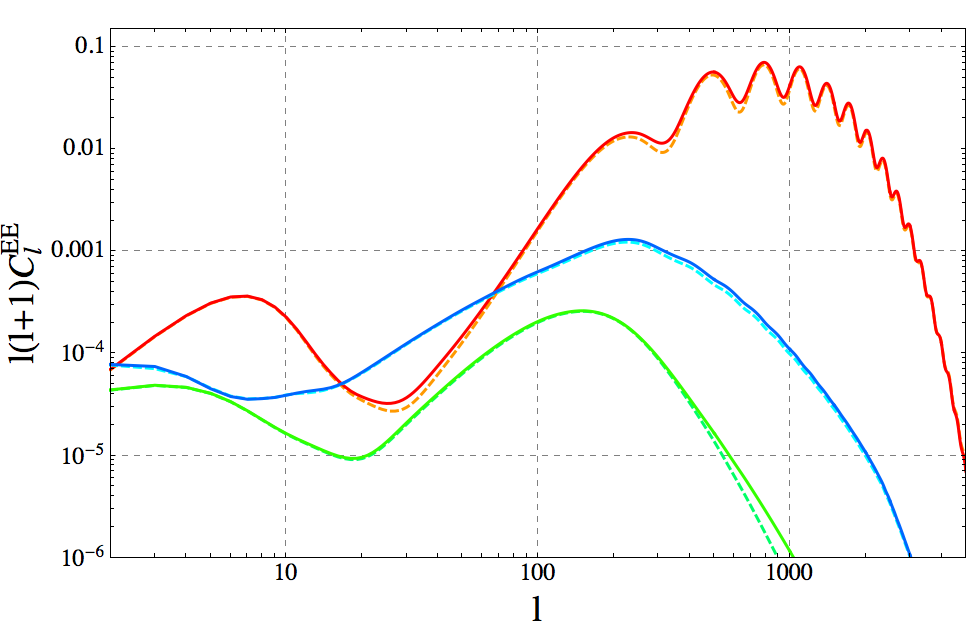}
\end{center}
\caption{The $C_\ell^{EE}$ spectrum decomposed into its scalar (top red/orange curves at $\ell = 100$), vector (bottom green curves at $\ell = 100$) and tensor (middle blue curves at $\ell = 100$) parts. Also here, procedures 2 and 3 differ much more significantly for the scalar and tensor modes than for the vector mode.}
\label{fig:partialEE}
\end{figure}

Let us also discuss the contributions of each type of perturbation, scalar, vector and tensor, to each CMB power spectrum. In Figs.~\ref{fig:partialTT}-\ref{fig:partialEE}, we show separately the power spectra  sourced only by the eigenvectors from tensor, vector and scalar UTCs. The color coding/line style is again common to all Figs.~\ref{fig:partialTT}-\ref{fig:partialEE} (though different than in Figs.~\ref{fig:totalTT}-\ref{fig:totalEE}), dashed and continuous lines corresponding to procedures 2 and 3 respectively, and red/orange to scalar perturbations, green to vector perturbations, and blue to tensor perturbations. For each case we plot the amplitudes obtained from procedures 2 and 3. This allows us to identify the contribution which is most affected by the more realistic treatment of the evolution around $t_\eq$. In Fig.~\ref{fig:partialTT} we see that the TT power spectrum is dominated by the scalar contribution over the entire $\ell$ range. Tensor contributions have a discrepancy of $\sim 10\%$ between procedure 3 versus procedure 2 (from $\ell > 250$), but we conclude that the $\sim 25\%$ difference in the total temperature power spectrum is mainly due to the discrepancy (of the same order) in the scalar contribution, which dominates completely over the vector and tensor contributions\footnote{Note that the vector contribution shows a signifcant discrepancy of $\mathcal{O}(100) \%$ for $\ell > 400$, but this is irrelevant since at small scales the vectors are really subdominant (even more than at low $\ell$) as compared to the scalar contribution. An analogous feature, i.e. a significant deviation of the vectors between methods 2 and 3 at very small scales, actually shows up in the rest of the channels, BB, TE and EE. However, again this is an irrelevant issue, since at those scales the vector contribution is always very subdominant versus either the tensor or the scalar contributions.}. In Fig.~\ref{fig:partialBB} we show the analogous decomposition but for the BB power spectrum. First of all, both tensor and vector contributions are relevant, essentially of the same order, though vectors are slightly dominating, except for the interval around $\ell \approx 15-60$, as well as for small scales, $\ell \gtrsim 350$. Procedures 2 and 3 do not differ significantly for either tensor or vector perturbations in this channel. The deviation between the amplitudes obtained with procedure 3 and with procedure 2 are indeed of similar order for vectors and tensors. The total discrepancy among the different procedures in the final BB spectrum over the full $\ell$ range (see previous comments about Fig.~\ref{fig:totalBB}, is due to a combination of both the tensor and vector contributions. We can say that this channel is the least affected by the choice of the procedure.

In Fig.\ \ref{fig:partialTE} we show the same decomposition for the TE channel. In this case both tensor and vector contributions have some few percent discrepancy when comparing the results of procedure 3 and procedure 2. But as in the case of of the TT channel, the final curve is completely dominated by the scalar contribution, which therefore is responsible for the final differences in amplitude of the total spectrum (which we reported before when discussing Fig.~\ref{fig:totalTE}).

Finally, in Fig.\ \ref{fig:partialEE}, we report the analogous decomposition for the EE channel. The difference in amplitude between procedure 3 and procedure 2, is mostly due to the scalar contribution for almost the entire $\ell$ range, except for the small interval $\ell \approx 15-50$ for which the tensors dominate. Like for the temperature, the discrepancies among the procedures are of the same order in the scalar and tensor contributions, and smaller for the vector mode. The differences in the final spectrum of this channel are due essentially to only the scalar contribution, since the latter dominates over the entire $\ell$-range. Note that the first bump is like in the inflationary case due to reionization, whereas for the rest of the spectrum the peaks are out of phase with respect to the inflationary signal, since defects produce isocurvature perturbations.

The result that the vector mode is not very sensitive to the chosen procedure might be related to fact that vector perturbations of the CMB do not oscillate. This can lead to a  more stable scalar product for vector modes. Note however, that the decay rate for vector perturbations is different for the two procedures. However, this shows up only in the highly suppressed tail at small scales, $\ell > 400$, and has no observational relevance. 

Note that all the power spectra shown have been obtained summing up the contributions from the first 200 eigenvectors of each type of perturbation (scalar, vector or tensor), i.e.~$C_\ell \equiv \sum_{n = 1}^{200} C_\ell^{(n)}$, with $C_\ell^{(n)}$ the contribution from the $n$th eigenvector. The convergence of the successive adition of contributions is indeed quite fast, verifying $\left|\sum_{n = 1}^{m} C_\ell^{(n)} - C_\ell\right| \ll 0.01C_\ell$ already for the first $m = 40$ eigenvectors, in almost every perturbation and channel (the exception being the BB polarization at small scales $\ell > 1000$, where the convegence is slower).

In all the results that we will present in the following we only use procedure 3, which for the large-$N$ model is more than 1\% accurate. Interestingly, the spectra obtained for the large-$N$ using procedure 3 are typically somewhat larger than those obtained from the standard procedure 2, typically of $\mathcal{O}(10)\%$, depending on the channel and the multipolar scale $\ell$. For the convenience of the reader and of workers in the field, all the spectra $C_\ell^{XY}$ are available on the homepage of the Geneva cosmology group: {\tt http://cosmology.unige.ch/research} under 'data products'.

Let us finally remark that, of course, when using the large-$N$ case as a template for monopoles or $O(4)$ textures, intrinsic differences of the order of 10\% have to be added to the error budget. Moreover, the case of cosmic strings is simply not well described by the large-$N$ scenario. Therefore, we cannot quantify how presently published CMB spectra from topological defects, particularly from the most relevant case of cosmic strings, will be affected by recalculating them using a similar treatment as the procedure 3 outlined here. Our present work, although based only on the large-$N$ case, suggests that it is possible that differences of $\mathcal{O}(10)\%$ might arise. Therefore, the accuracy of previously calculated CMB power spectra from topological defects should be taken with caution, at least until an equivalent methodology to procedure 3 is employed. We consider this observation as a first important result of our present work.

\subsection{The correlation functions}

\begin{figure}[t]
\begin{center}
\includegraphics[width=0.75\linewidth]{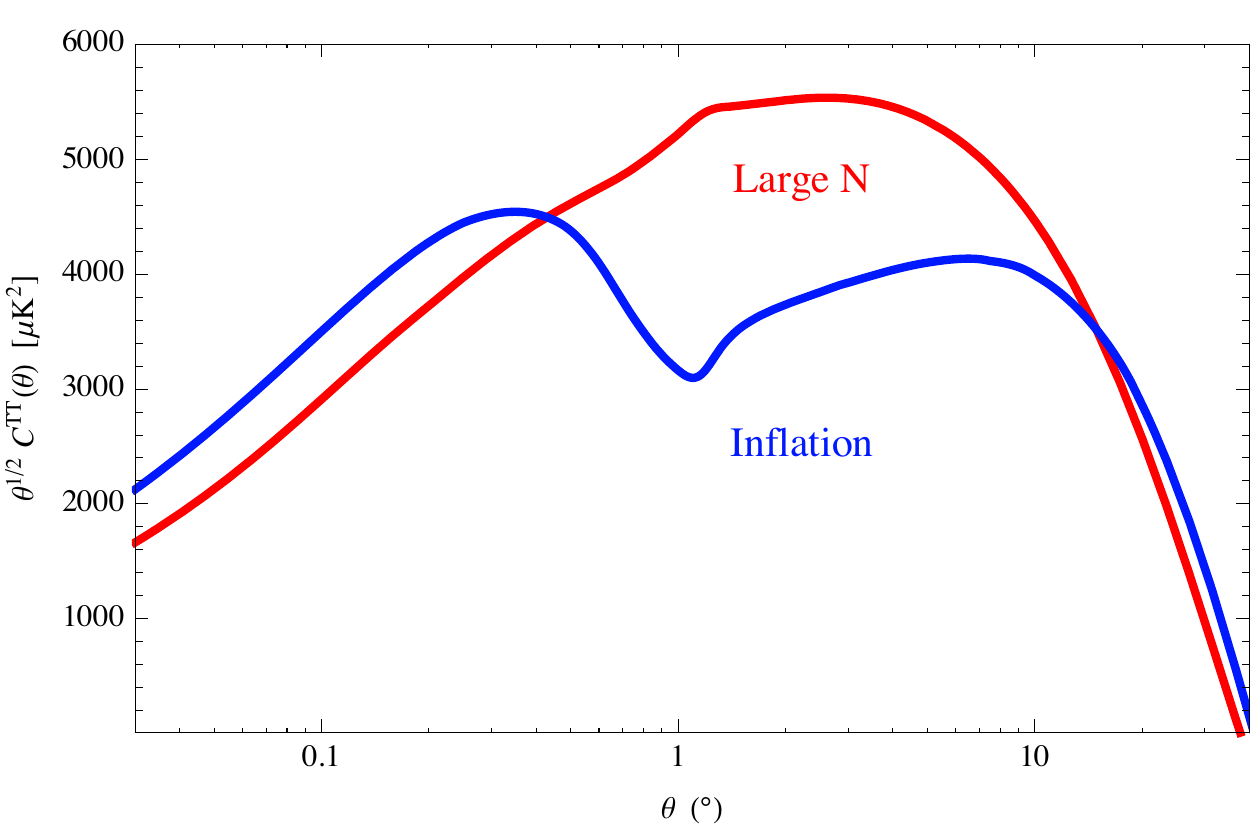}\\
\includegraphics[width=0.75\linewidth]{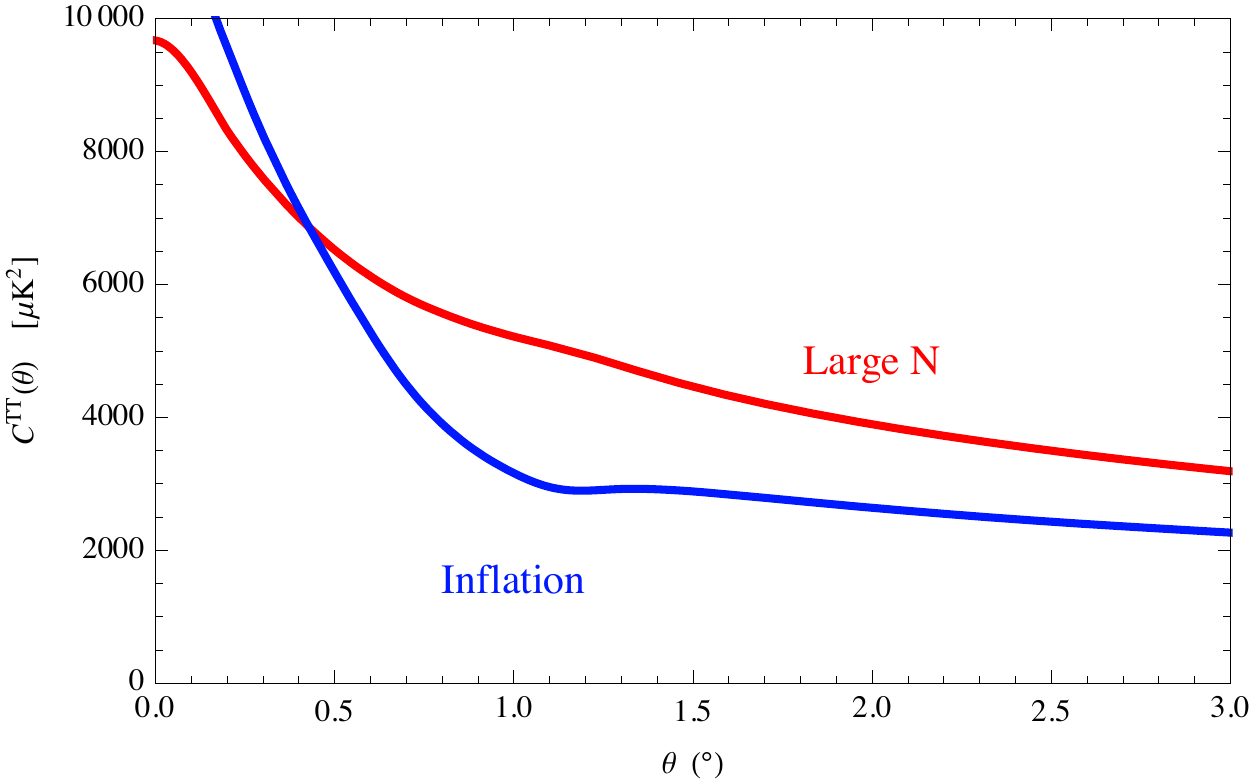}\\
\includegraphics[width=0.75\linewidth]{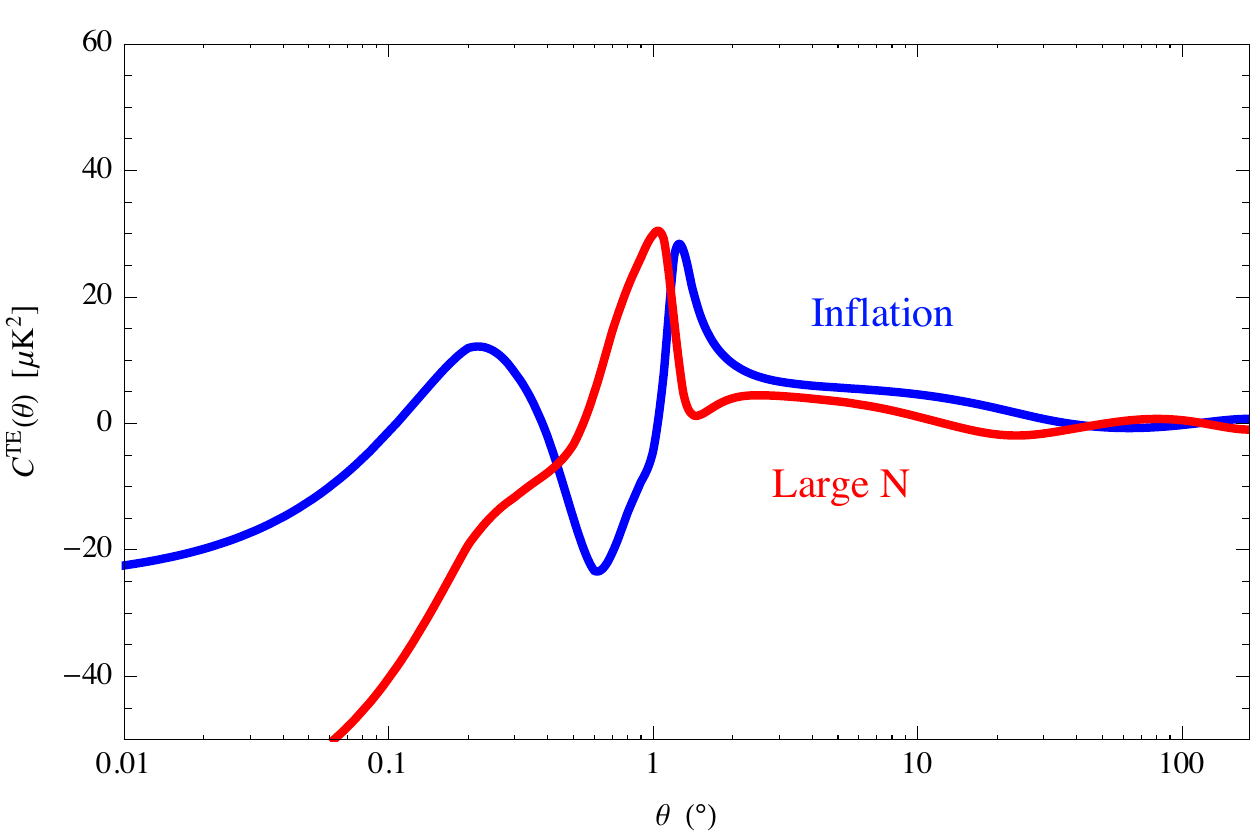}\\
\includegraphics[width=0.75\linewidth]{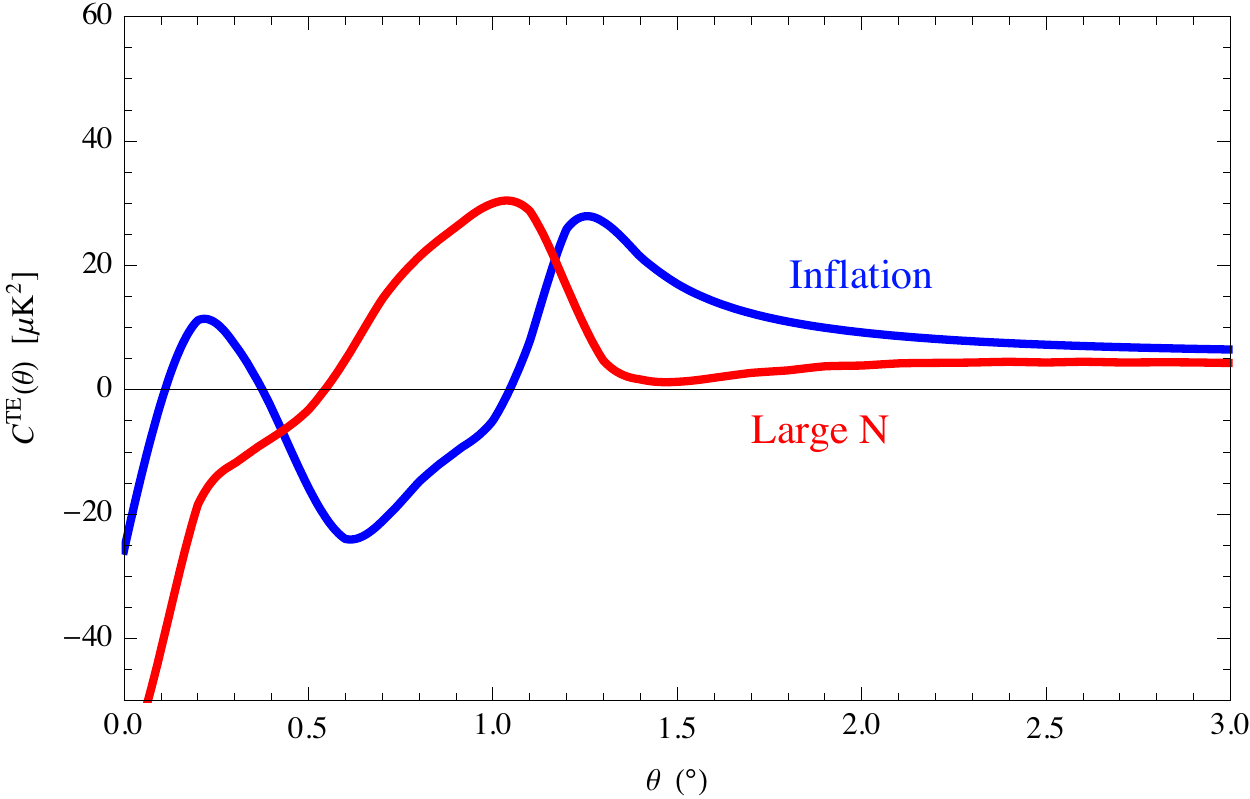}
\end{center}
\caption{The correlation functions from large-$N$ and from inflation. Clearly, the main difference is at angular scales $\theta<3^o$. Note the characteristic sign difference of the TE correlation functions. The large-$N$ contribution is normalized such that $C_\ell^{TT \;{\rm large N}}=0.1C_\ell^{TT\; {\rm inf}}$ at $\ell=10$, and then it is multiplied by a factor 50 for visibility.} 
\label{fig:corfsT}
\end{figure}

The correlation functions of the CMB temperature anisotropies and polarization are given by
\be
\xi^{XY}(\theta) = \frac{1}{4\pi}\sum_\ell (2\ell +1)C_{\ell}^{XY}P_\ell(\cos\theta)\,,
\ee
where $X$ and $Y$ denote as before $T$, $E$ and $B$ and $P_\ell$ is the Legendre polynomial of order $\ell$.
In principle, the power spectrum and the correlation function contain exactly the same information. However, the form of certain correlation functions from defect sources have very characteristic shapes, which can make it easier to distinguish them from inflation, than by looking at the power spectrum. A similar situation is known from the acoustic peaks in the matter power spectrum which are easier to see in the correlation function. 
In Fig.~\ref{fig:corfsT} we show the $TT$ and $TE$ correlation functions. The acoustic peak at $\theta\simeq 1^o$ is a pronounced minimum and kink in the inflationary TT correlation function. For the large-$N$ model there is only a slight kink. Also, in the TE correlation function the acoustic peaks show up as a pronounced double-maximum well separated by a minimum, while in the large-$N$ TE correlation function the first maximum and the minimum are entirely missing. This is a consequence of causality as has been pointed out already in~\cite{Spergel:1997vq} and tested with a toy model in~\cite{Scodeller:2009iu}.

As we have pointed out in~\cite{GarciaBellido:2010if}, the so called  'local' polarization correlation functions (for an introduction see~\cite{Durrer:2008aa}) are especially useful. Let us repeat their definition here. The usual polarization correlation functions and of course all the CMB power spectra are non-local, i.e. they require in principle information from the entire sky. Local polarization correlation functions can be defined as follows.

Polarization of the CMB is described as 
a rank-2 tensor field ${\cal P}_{ab}$ on the sphere.
It is usually decomposed into Stokes parameters,
${\cal P}_{ab}=(I\si_{ab}^{(0)} +U\si_{ab}^{(1)} +V\si_{ab}^{(2)}+
Q\si_{ab}^{(3)})/2=I\delta_{ab}/2 +P_{ab}$,
where $\si^{(j)}$ are the Pauli matrices and $\si_{ab}^{(0)} = \delta_{ab}/2$~\cite{Durrer:2008aa}.
The variable $I$ corresponds to the intensity of the radiation and contains the 
temperature anisotropies. As the relevant scattering process at late times -- Thomson scattering -- does not induce circular polarization we expect $V=0$ for the CMB polarization, and hence $P_{ab}$ to be real.
For a given direction $\bn$ we define the orthonormal frame 
$(\bfe_1,\bfe_2,\bn)$ and the circular polarization vectors 
$\bfe_\pm = \frac{1}{\sqrt{2}}\left(\bfe_1\pm i\bfe_2\right)$ as before. This  allows us
to introduce the components $P_{\pm\pm}= 2\bfe_\pm^a \bfe_\pm^bP_{ab} 
= Q \pm i U$ and $P_{+-} \propto V =0$.  The second derivatives
of this polarization tensor are related to the {\em local} $\tilde{E}$- and 
$\tilde{B}$-polarizations,
\bea\nonumber
\nabla_-\!\nabla_-P_{++} + \nabla_+\!\nabla_+P_{--} &\!=\!& 2\nabla_a\!\nabla_bP_{ab}
 \equiv \tilde E\,, \\ \nonumber
\nabla_-\!\nabla_-P_{++} - \nabla_+\!\nabla_+P_{--} &\!=\!& 2\ep_{cd}\ep_{ab}
\nabla_c\!\nabla_aP_{bd} \equiv \tilde B \,. \hspace{2mm}
\eea

\begin{figure}
\begin{center}
\includegraphics[width=0.75\linewidth]{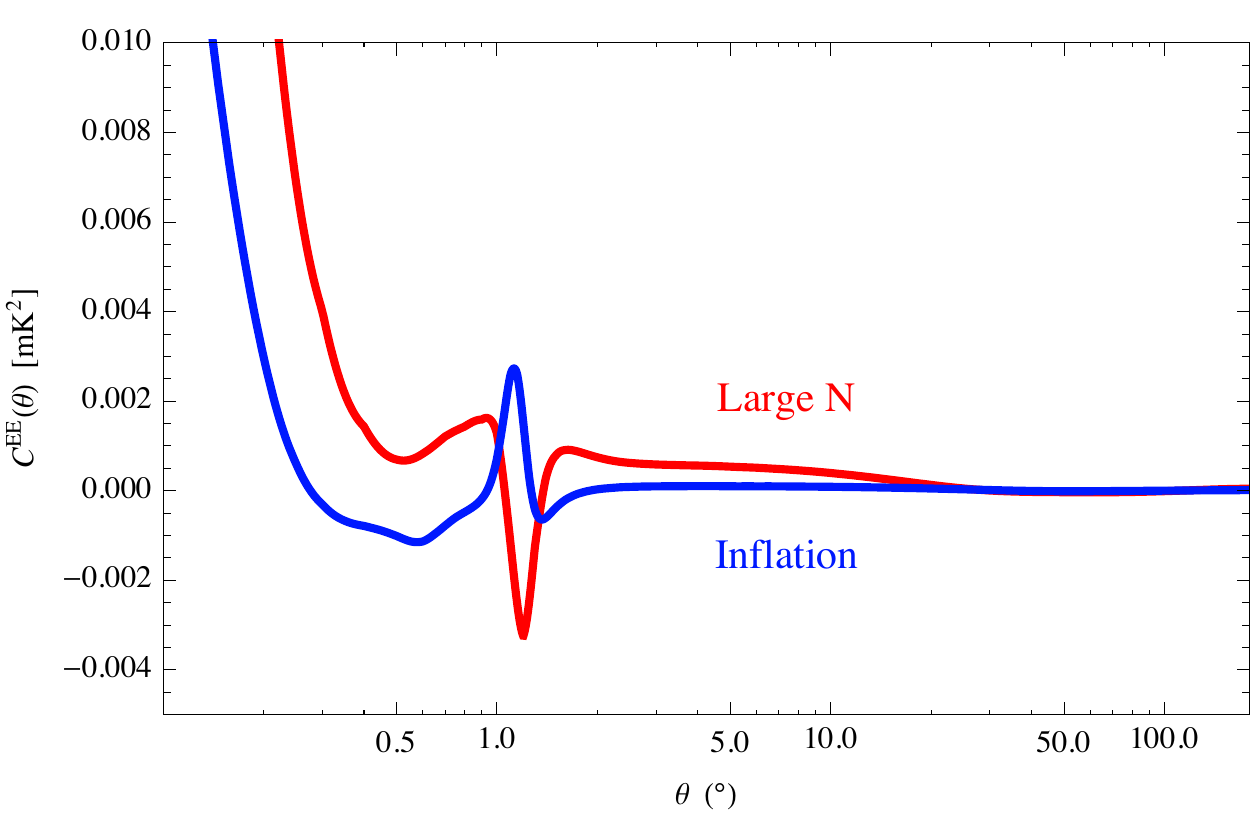}\\
\includegraphics[width=0.75\linewidth]{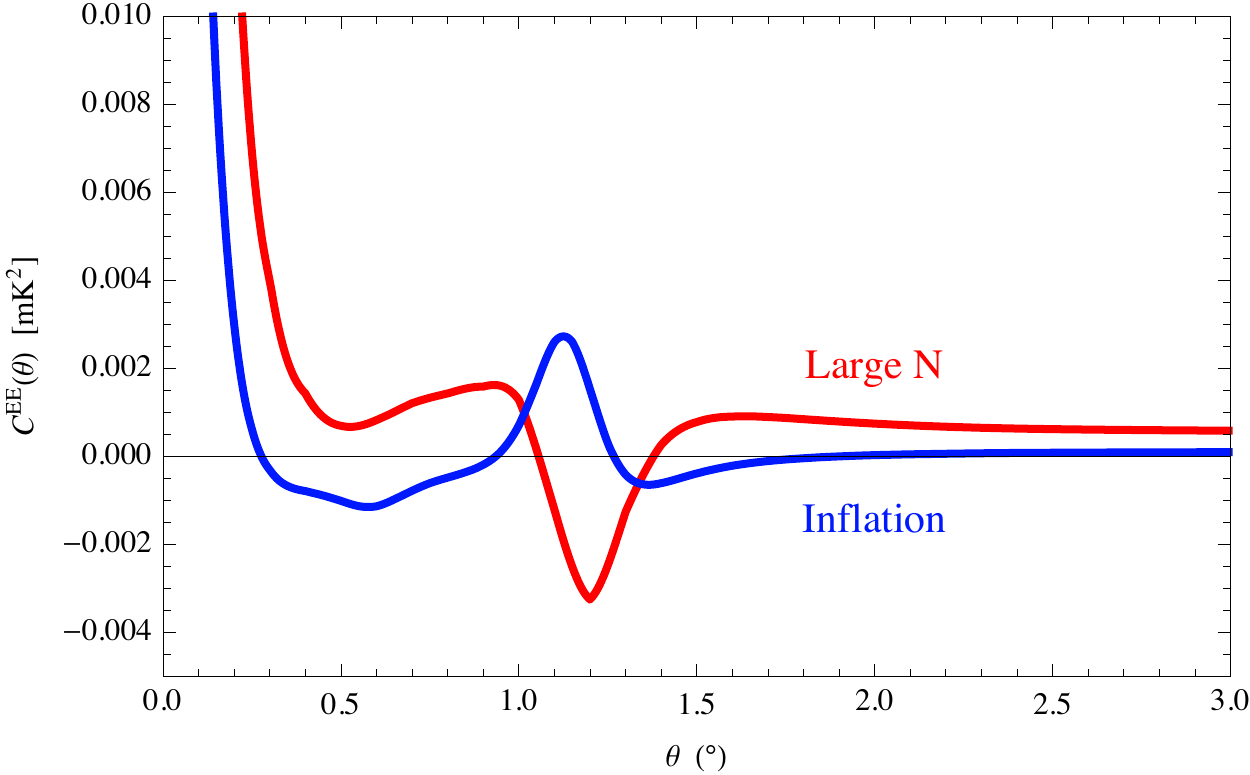}\\
\includegraphics[width=0.75\linewidth]{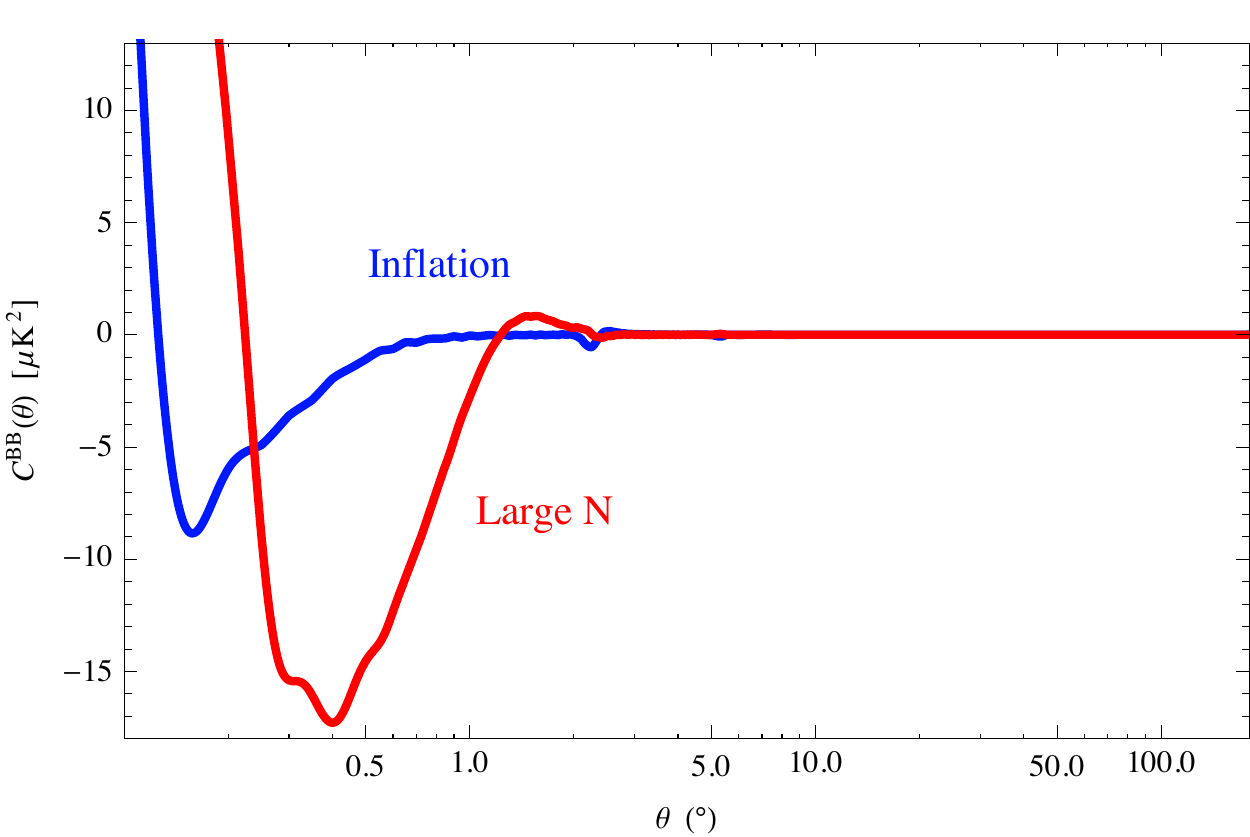}\\
\includegraphics[width=0.75\linewidth]{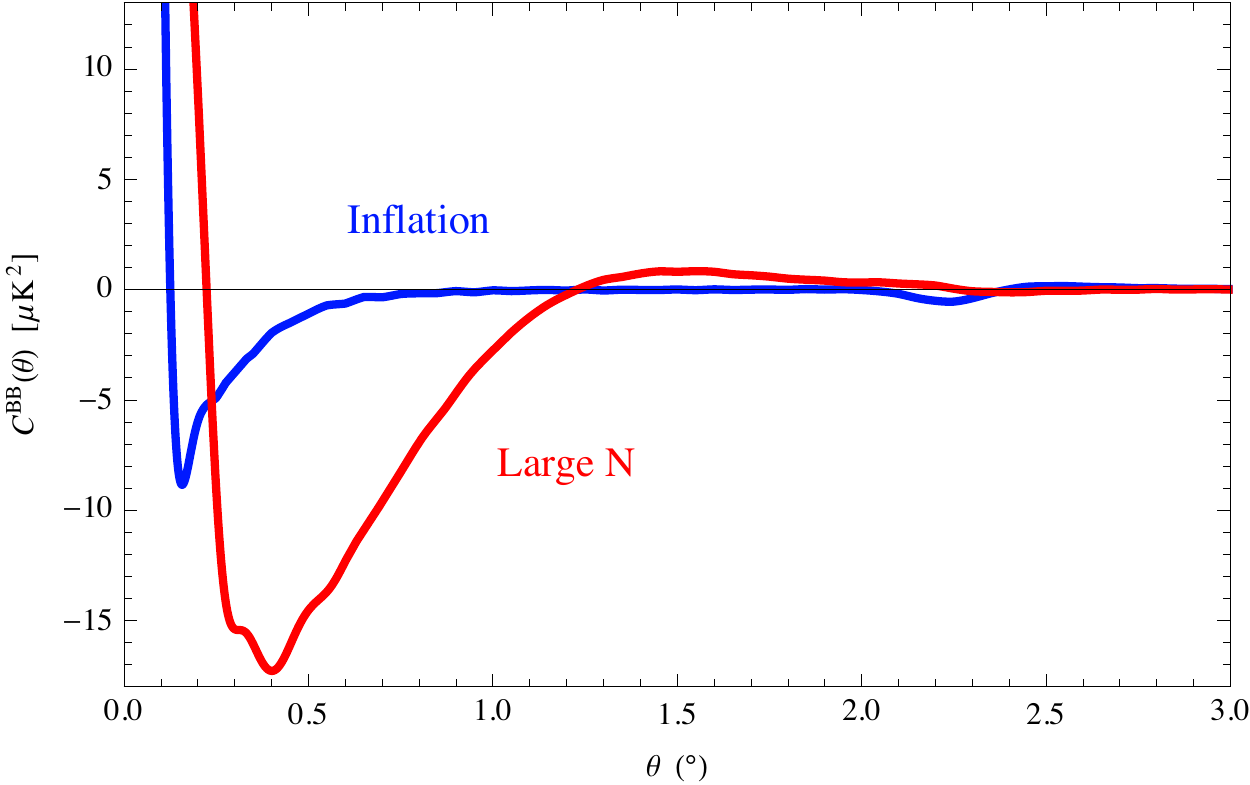}

\end{center}
\caption{The local correlation functions for E- and B-polarization from large-$N$ and from inflation.  The B-correlations on angular scales $\theta<1.5^o$ are very different from the inflationary ones. Also the E-polarizations have opposite signs at about $1^o$ degree. Again, the large-$N$ contribution has been multiplied by 50 in order to compare with inflation. The $BB$ contribution due to gravitational waves from inflation is normalized such that $C_\ell^{BB \;{\rm inf}}=0.1C_\ell^{TT\; {\rm inf}}$ at $\ell=10$.}
\label{fig:corfEB}
\end{figure}

Here $ \nabla_{\pm}$  are the derivatives in the directions 
$ \bfe_{\pm} $ and
$\ep_{cd}$ is the 2-dimensional totally anti-symmetric tensor.
These functions are defined {\em locally}. The usual $E$- and $B$-modes 
can be obtained by applying the inverse of the  Laplacian, $\nabla^2 = \nabla_+ \nabla_- + \nabla_- \nabla_+$  to the local $\tilde E$- and 
$\tilde B$-polarizations. Such inversions of differential operators depend on boundary 
conditions which can affect the result for local observations.
The $\tilde{B}$ (and $\tilde{E}$) correlation functions, $\xi^{\tilde B}(\theta) \equiv  \langle 
\tilde B(\bn)\tilde B(\bn')\rangle_{\bn\cdot\bn' =\cos\theta}$, and
$\xi^{\tilde E}(\theta) \equiv  \langle 
\tilde E(\bn)\tilde E(\bn')\rangle_{\bn\cdot\bn' =\cos\theta}$
are measurable locally. They are related to the  power spectrum
by~\cite{Durrer:2008aa}
\be\label{CBL}
\xi^{\tilde B,\, \tilde E}(\theta) =\frac{1}{4\pi}\sum_{\ell=2}^\infty 
 \frac{(\ell+2)!}{(\ell-2)!}(2\ell+1)P_\ell(\cos\theta)
C_{\ell}^{B,\, E} \,.
\ee
The additional factor $n_\ell = (\ell+2)!/(\ell-2)! \sim \ell^4$ enhances the power on small scales where the defect polarization is enhanced with respect to the inflationary one.

In Fig. \ref{fig:corfEB} we show the local correlation functions of $E$ and $B$ polarization for the large-$N$ model and we compare them with the corresponding ones from inflation. Clearly, at small angles, $\theta \lesssim 3^o$ their shape is very different. Around $\theta=1.2^o$, inflationary E-polarisation has a maximum while E-polarization from large-$N$ has a minimum. Also the most significant structure of B-polarisation from the large-$N$ model is around $\theta \sim 0.3^o$ to $1^o$, where the inflationary signal is already small. In a previous paper~\cite{GarciaBellido:2010if} we have exploited this fact, which is even more pronounced for topological defects like cosmic strings or global monopoles and texture,  to predict limits on a defect contribution from future B-polarisation measurements.
 
In the next section we shall use the shape of the power spectrum to identify a possible small large-$N$ component in the CMB sky.

\section{Comparison to present and future experiments}\label{s:exp}

In this section we want to estimate the signal-to-noise ratio of the different channels for three different types of experiments: 1) Planck, 2) a COrE-like experiment, 3) PRISM and 4) an optimal polarization experiment which is cosmic variance limited out to $\ell=10000$.
We assume that the true signal is of the form
\be\label{mix}
C_\ell = (1-\ep^2)C_\ell^{\rm inf} + \ep^2 C_\ell^{\rm large N} \,,
\ee 
and we want to study how well a given experiment can constrain the amplitude $\ep$.
In (\ref{mix}) we normalize $C_{20}$ to the observed value,
$$C_{20}^{\rm inf} = C_{20}^{\rm large N} =915.92 \, (\mu K)^2\frac{2\pi}{20\times21}\,.$$

We assume that the covariance matrix is diagonal in $\ell$-space, i.e. that not only the signal but also the noise is statistically isotropic. For Gaussian $C_\ell$'s we then have
\bea
&&\hspace*{-0.5cm} {\rm Cov}[C_\ell^{XY},C_{\ell'}^{VW}] = \frac{\left({\cal C}_\ell^{XV}{\cal C}_\ell^{YW} +
{\cal C}_\ell^{XW}{\cal C}_\ell^{YV}\right)}{f_{\rm sky}(2\ell+1)}\de_{\ell\ell'} \,,\\  &&
 {\cal C}_\ell^{XV} = (C_\ell^{XV}+N_\ell^{XV})\exp[\ell(\ell+1)/\ell_s^2]\,.
\eea
Here $f_{\rm sky}$ is the fraction of the observed sky, $N_\ell$ is the noise spectrum of the experiment and $\ell_s$ is a smoothing scale which must be larger than the resolution of the experiment,
$$
 \ell_s<\ell_r = \frac{\sqrt{8\ln(2)}}{\theta_{\rm FWHM}}\,,  \quad (\theta_{\rm FWHM}=\mbox{beam width})\,.
$$

The noise spectrum is  assumed to be white noise characterized by an amplitude
$\De_{P,{\rm eff}}$ for polarization and $\De_T$ for temperature noise,
$$
N^X_\ell = \frac{ \ell(\ell + 1)}{2\pi}\exp\left[ \frac{\ell (\ell + 1)}{\ell_r^2} \right] \De_X^2 \,.
$$

\begin{table}[h!]
\centering
\begin{tabular}{|c|c|c|c|c|}
\hline
  & Planck & COrE & PRISM & CVL \\
\hline
$\theta_{\rm FWHM}$  & 7.2 & 4.7  & 2.3 & 1.0 \\ 
$\De_{P,{\rm eff}}$  & 23.4 & 2.05  & 1.43 & 0 \\ 
$\De_T$  & 14.5 & 1.19  & 1.01 & 0 \\ 
Ref. & PLA \cite {wPlanck}& \cite{wCore} & ~\cite{wPrism} & noiseless \\
\hline
\end{tabular}
\caption{The FWHM beam width, in arcmin, and the noise level of the different CMB experiments considered here, in units of  $\mu$K$\cdot$arcmin.}
\label{tab:1}
\end{table}

We shall consider the experiments listed in Table~\ref{tab:1}. When we consider a signal to which only the defects contribute, we can simply calculate the signal to noise from the large-$N$ contribution either with the correlation function or the $C_\ell$'s via
\be
\frac{S}{N} =\sum_{\ell=2}^{\ell_{\max}}\frac{C_\ell}{N_\ell}\,.
\ee

As we have seen in Ref.~\cite{GarciaBellido:2010if}, for the B-mode, most of the large-$N$ or defect signal comes from the correlation function at small angles, 
$\theta<1^o$. 

However, when both, large-$N$ and inflation contribute to the signal, $S/N>1$ is not sufficient to detect the signal, but we must also be able to distinguish it from the inflationary signal. To quantify this we use the Fisher matrix technique. We consider a signal to which not only the large-$N$ (or defects) contribute but there is also an inflationary contribution. We split the signal as
\be
{\sf C}_\ell = \al C_\ell^i +\beta C_\ell^d + N_\ell \,,
\ee 
where $C_\ell^{i,d}$ denotes the inflationary (i) and large-$N$ defect (d) signal, normalized such that they are equal at $\ell=20$,
$$
C_{20}^i = C_{20}^d =915.92 \, (\mu K)^2\frac{2\pi}{20\times21} \,.
$$
Here $C_\ell$ can mean the temperature, $C^{TT}_\ell$, the E-polarization, $C^{EE}_\ell$ or the temperature-polarization cross-correlation,
$C^{TE}_\ell$.

\begin{table}
\centering
\begin{tabular}{|c|c|c|c|c|}
\hline
$\delta\beta$ & Planck & COrE & PRISM & CVL \\
\hline
TT & $1.4\cdot10^{-2}$ & $3.5\cdot10^{-3}$ & $8\cdot10^{-4}$ & $3\cdot10^{-4}$  \\
\hline
EE & $6\cdot10^{-3}$ & $4.1\cdot10^{-3}$ & $4\cdot10^{-3}$ & $3.7\cdot10^{-3}$  \\
\hline
TE & $4\cdot10^{-4}$ & $3\cdot10^{-5}$ & $2.5\cdot10^{-5}$ & $2.4\cdot10^{-5}$  \\
\hline
\end{tabular}
\caption{The TT, EE and TE errors on $\beta$, for Planck, a COrE-like experiment, PRISM  and a CVL experiment. We set
$f_{\rm sky}=0.7$ in all cases.
\label{tab:betaTT}}
\end{table}

Studying how well we can measure different parameters in a signal depending on several parameters, is best done with the Fisher matrix technique. The resulting limits assume that the errors on the parameters are Gaussian, which is often not true, but nevertheless, if the errors are small enough it is usually a good approximation. For $C_\ell$'s depending on a series of parameters $\la_i$ the Fisher matrix is determined by
\be
F_{ij} = \sum_{\ell}\frac{\dd {\sf C}_\ell}{\dd\la_i}\frac{\dd {\sf C}_\ell}{\dd\la_j}{\rm Cov}_\ell^{-1} \,,
\ee
where we have already used that the covariance matrix is diagonal, 
${\rm Cov}_{\ell\ell'} =\de_{\ell\ell'}{\rm Cov}_\ell$. We are only interested in the two parameters $\la_1 \equiv \al$ and $\la_2 \equiv \beta$. We will fix all other cosmological parameters to the values measured by the Planck collaboration. In the light of the Planck results, the presence of defects (parametrized by their fractional contribution $f_{10}$ at multipole $\ell = 10$) is actually very weakly correlated with the standard cosmological parameter values~\cite{Ade:2013xla}, the latter being affected only in their third decimal digit by the inclusion of $f_{10}$ in the Monte Carlo Markov Chain analysis. Therefore, there is no danger of a degeneracy and we can fix the standard cosmological parameters to the values measured by Planck in the absence of cosmic defects. The Fisher matrix is obtained as
\bea
F_{\al\al} &= \sum_{\ell} (C_\ell^i)^2\frac{2\ell+1}{({\sf C}_\ell)^2}
\frac{f_{\rm sky}}{2}  &= F_{11} \,, \\
F_{\beta\beta} &= \sum_{\ell} (C_\ell^d)^2\frac{2\ell+1}{({\sf C}_\ell)^2}
\frac{f_{\rm sky}}{2}  &= F_{22} \,, \\
F_{\al\beta} &= \sum_{\ell} C_\ell^iC_\ell^d\frac{2\ell+1}{({\sf C}_\ell)^2}
\frac{f_{\rm sky}}{2}  &= F_{12} = F_{21} \,.
\eea
and its inverse is given by
\be
\left(F^{-1}\right)_{ij} = \frac{1}{\det F}\left(\begin{array}{cc}
    F_{\beta\beta} & -F_{\al\beta}\\ -F_{\al\beta} & F_{\al\al} \end{array}\right)
\ee
The marginalized error on the parameter $\beta =\beta_0\pm\de\beta$ is now simply given by  (see \cite{Durrer:2008aa}, p231ff)
\be
(\de\beta)^2 = (F^{-1})_{22} = \frac{F_{\al\al}}{F_{\al\al}F_{\beta\beta} - (F_{\al\beta})^2}  \,.
\ee
We have to compute $\de\beta$ for some fiducial values $(\al_0,\beta_0)$. We found that the result is nearly independent of $\beta_0$ for $\beta_0<0.1$, hence we can set $\beta_0=0$. The value of $\de\beta$ together with the normalization of $C^d_\ell$ then determines the VEV of the large-$N$ field, or equivalently the combination $Gv^2/N$. See table~\ref{tab:EpsilonTT}.

\begin{table}[t]
\centering
\begin{tabular}{|c|c|c|c|c|}
\hline
$Gv^2/N$ & Planck & COrE & PRISM & CVL \\
\hline
TT & $5.2\cdot10^{-7}$ & $2.7\cdot10^{-7}$ & $1.3\cdot10^{-7}$ & $7.8\cdot10^{-8}$  \\
\hline
EE & $3.5\cdot10^{-7}$ & $2.9\cdot10^{-7}$ & $2.9\cdot10^{-7}$ & $2.7\cdot10^{-7}$  \\
\hline
TE & $9.0\cdot10^{-8}$ & $2.5\cdot10^{-8}$ & $2.3\cdot10^{-8}$ & $2.2\cdot10^{-8}$  \\
\hline
\end{tabular}
\caption{The upper bounds on $Gv^2/N$ from the TT, EE and TE errors on $\beta$, for Planck, COrE-like experiment, PRISM, and a CVL experiments. We set $f_{\rm sky}=0.7$ in all cases.
\label{tab:EpsilonTT}}
\end{table}

Interestingly, the TE-correlation gives the best constraints, even better than those from B-polarization for Planck~\cite{GarciaBellido:2010if}. Furthermore, there is virtually no additional gain when going from an experiment like COrE or PRISM to a cosmic variance limited experiment. This means that already for PRISM (or COrE) the dominant contribution to the uncertainly comes from cosmic variance and cannot be improved by better experiment technology. In this case, B-polarisation limits are much better than those from T and E-polarization. In Ref.~\cite{GarciaBellido:2010if} we have shown that a cosmic variance limited B-polarisation experiment can detect a large-$N$ signal down to $Gv^2/N = 1.4\times 10^{-10}$ which is more than two orders of magnitude better than the best limit we can achieve from T and E signals.
This confirms our claim of Ref.~\cite{GarciaBellido:2010if}, that B-polarisation is a very sensitive probe for cosmic defects, here in the case of large-$N$.

\section{Discussion and Conclusions}\label{s:con}

In this paper we have fully calculated the imprint on the CMB temperature anisotropy and polarization from  large-$N$ scaling seeds. This model has the advantage that the source term can be computed analytically. In addition, it is a good approximation for global monopoles, global textures and for more than four coupled global scalar fields, which in 3+1 dimension do not give raise to topological defects (but yet produce non-topological gradient field configurations, so called non-topological defects). We have found that the breaking of scale invariance which happens at  the transition from radiation to matter is important and leads to an imprint on the CMB power spectra. Taking it into account by a simple interpolation from the radiation dominated to the matter dominated source, at the level of the eigenvectors, leads to errors of the order of up to 25\%. Only after using of the order of 10 interpolation steps can we trust our result to be accurate at the 1\% level. 
 
We have found the time-dependence of a universal (scale independent) interpolation function that one can use to weight correctly the MD and RD correlators at every moment (within one of the chosen subintervals). Besides, we have proposed a prescription, using the previous interpolation function, for introducing active sources in CMB codes accounting accurately for the contribution of such sources around $t_{\rm eq}$. It would be very interesting to test this procedure also on topological defects like cosmic strings.

Finally, we have investigated how well such a component can be detected in the CMB. This can be cast in terms of upper limits for $Gv^2/N$. We have found that for experiments with considerable noise, like Planck the TE correlation is the most sensitive channel, while for a cosmic variance limited or very low noise experiment like PRISM the B-polarisation channel is a more than two orders of magnitude more sensitive probe. 

We have made available our final CMB spectra at {\tt http://cosmology.unige.ch/research} (under 'data products') for the convenience of the reader.

\acknowledgments{We are very grateful to Mark Hindmarsh, Joanes Lizarraga and Jon Urrestilla, for critical discussions and comments on the draft. We also acknowledge helpful discussions with Jos\'e Juan Blanco Pillado and Martin Bucher. This work is supported by the Swiss National Science Foundation. Some numerical calculations were run on the Andromeda cluster of the University of Geneva. The authors also acknowledge financial support from the Madrid Regional Government (CAM) under the program HEPHACOS S2009/ESP-1473-02, from the Spanish MICINN under grant AYA2009-13936-C06-06 and Consolider-Ingenio 2010 PAU (CSD2007-00060), and from the MINECO, Centro de Excelencia Severo Ochoa Programme, under grant SEV-2012-0249, as well as from the European Union Marie Curie Initial Training Network UNILHC PITN-GA-2009-237920.}

\begin{widetext}
\appendix
\section{The unequal time correlators}\label{ap:utc}

\subsection*{Equal time correlators and asymptotic behaviour of scalar sources}

As discussed in section \ref{s:formalism} we need to calculate the unequal-time two point functions of the metric perturbations
in order to be able to compute the CMB power spectra.
Here we present analytic expressions for the unequal time correlators of the scalar field energy momentum tensor and we discuss their asymptotic behavior, from \cite{mkthesis}.
The energy density, $f_\rho = [(\dot\beta)^2 + \nabla\beta)^2]/2$ can be expressed in terms of the exact solution given in (\ref{e:beta}).
To simplify the notation, we set
\be
\chi(x)\equiv \frac{J_\nu(x)}{x^\nu} \, , \quad \vph(x) \equiv \left(\frac{3}{2}\chi(x)
	-\frac{J_{\nu+1}(x)}{x^{\nu-1}}\right)\,.
\ee
where the form of $\vph$ arises from $\dot{\beta}$ in the energy momentum tensor, with the derivative of $J_\nu(x)$ re-expressed with the help of the usual relation for Bessel functions, $d/dx [J_\nu(x)/x^\nu] = - J_{\nu+1}(x)/x^\nu$.

We also introduce  the dimensionless
variables $x\equiv qt$ and $y \equiv kt$. Furthermore, the products in real space become convolutions in Fourier space,
and it is useful to introduce for expressions like $g(x) h(|\by-\bx|)$ the compact notation 
 $g(x) h(|\by-\bx|)\equiv (gh)$. We also use that the random variables $\beta^a$ are Gaussian so that products like $\langle \beta^a(k)\beta^b(p)\beta^c(k')\beta^d(p')\rangle$ can be reduced via Wick's theorem to sums over products of two-point expectations, which in turn are given by Eq.\ (\ref{e:coreta*}). With this we obtain
\bea
\left\lan|f_\rho^2|\right\ran (\by,t) &=&
\frac{(2\pi)^3}{2N\NN^2 t} \int d^3\!x\, \left\{(\vph\vph)^2
+ [\bx(\by-\bx)]^2(\chi\chi)^2-2 [\bx(\by-\bx)] (\vph\vph)(\chi\chi)\right\} \\
&=& \frac{(2\pi)^4}{2N\NN^2 t} \int dx\,d\mu\,x^2 \left\{(\vph\vph)^2
+ [xy\mu-x^2]^2 (\chi\chi)^2-2 [xy\mu-x^2] (\vph\vph)(\chi\chi)\right\}.
\eea
where $\NN = 16/15$ for RD ($\nu = 2$) or $\NN = 128/2835$ for MD ($\nu = 3$). In the last equation we have performed the integration over one
angular variable and introduced $\mu=\cos \phi$. The pressure $f_p$ contains  the same terms, only the pre-factors differ.
The pre-factor of the $(\chi\chi)^2$ term is $1/9$ and the one of $(\vph\vph)(\chi\chi)$ is $2/3$.

From the above expressions it is clear that $f_\rho$ and
$f_p$ behave like white noise on super horizon scales.
Numerically we have found
\bea
\left\lan |f_\rho|^2\right\ran(y=0,t) &=& \frac{(2\pi)^4}{N t\NN^2} \left\{
	\begin{array}{ll} 1.72\cdot 10^{-2} & \mbox{for $\nu=2$}\\
		3.34\cdot 10^{-5} & \mbox{for $\nu=3$} \end{array} \right., \\
\left\lan |f_p|^2\right\ran(y=0,t) &=& \frac{(2\pi)^4}{N t\NN^2} \left\{
	\begin{array}{ll} 1.96\cdot 10^{-3} & \mbox{for $\nu=2$}\\
		2.61\cdot 10^{-6} & \mbox{for $\nu=3$} \end{array} \right. .
\eea

In the limit $y\gg 1$ they decay like
\bea
\left\lan |f_\rho|^2\right\ran(y\gg 1,t) &\sim&  y^{1-2\nu} t^{-1} ,\\
\left\lan |f_p|^2\right\ran(y\gg 1,t) &\sim&  y^{1-2\nu} t^{-1} .
\eea

$f_v$ is calculated analogously to $f_\rho$ and $f_p$:
\bea
f_v(\bk,t) &=& - i \frac{k_j}{k^2} \left(\dot{\b} \b_{,j}\right)(\bk,t) \\
	&=& - A t^2 \int d^3\!q \frac{\bk(\bk-\bq)}{k^2} \vph(qt) 
		\chi(|\bk-\bq|t)\b_\tin(\bq) \b_\tin(\bk-\bq).
\eea

Using the same dimensionless variables and the same
notation as above, we find for the equal-time correlator of $f_v$,
\bea
\left\lan |f_v|^2\right\ran(\by,t) &=& \frac{(2\pi)^3 t}{N\NN^2 y^4} \int d^3\!x
	\left\{\left[\by(\by-\bx)\right]^2(\chi\vph)^2 + \left[\by(\by-\bx)\right]
	\left[\by\bx\right](\chi\vph)(\vph\chi)\right\}\\
&=& \frac{(2\pi)^4 t}{N\NN^2 y^4} \int \!dx d\mu
	x^2 \left\{\left[y^2-x y \mu\right]^2 (\chi\vph)^2 + \left[y^2-x y \mu\right]
	\left[x y \mu\right](\chi\vph)(\vph\chi)\right\} \, .
\eea

A lengthy expansion around $y=0$ shows that the integrand vanishes
up to $y^3$, and that this term vanishes upon integration over $\mu$.
Therefore, $f_v$ does  not diverge for $y\rightarrow 0$. We obtain the finite result 
\be
\left\lan |f_v|^2\right\ran(y=0,t) = \frac{(2\pi)^4 t}{N\NN^2} \left\{
	\begin{array}{ll} 1.96\cdot 10^{-3} & \mbox{for $\nu=2$}\\
		2.61\cdot 10^{-6} & \mbox{for $\nu=3$.} \end{array} \right.
\ee
In the limit $y\gg 1$, $\left\lan |f_v|^2\right\ran$ decays like
\be
\left\lan |f_v|^2\right\ran(y\gg 1,t) \sim y^{-1-2\nu} t \, .
\ee

For $f_\pi$ we find
\bea
f_\pi(\bk,t) &=& -\frac{3}{2} \frac{k_i k_j}{k^4}
	\left(\b_{,i}\b_{,j}-\frac{1}{3} \delta_{ij} \left(\nabla\b\right)^2\right)\\
&=& \frac{3 A t^3}{2} \int d^3\!q \left[(\bk\bq)(k^2-\bk\bq)
	-\frac{1}{3} k^2 (\bk\bq-q^2)\right] \chi(qt) \chi(|\bk-\bq|t)
	\b_\tin(\bq) \b_\tin(\bk-\bq)\, .
\eea
The resulting equal time correlator is given by
\be
\left\lan f_\pi(\bk,t) f_\pi^*(\bk',t)\right\ran = \frac{9 (2\pi)^4 t^3}{2 N\NN^2} \int \!dx d\mu
	x^2 \frac{\left[xy\mu-x^2\mu^2+\frac{1}{3} \left(x^2-xy\mu\right)\right]^2}{y^4}
	(\chi\chi)^2  \, .
\ee

Clearly $f_\pi$  diverges for $y\rightarrow 0$ and we find
easily that
\bea
\left\lan |f_\pi|^2\right\ran(y\rightarrow 0,t) &=&
	\frac{9 (2\pi)^4 t^3}{2 N\NN^2 y^4} 
	\int_{-1}^1d\mu \left(\frac{1}{3}-\mu^2\right)^2
	\int_0^\infty dx x^6 \chi(x)^4 \nonumber \\
&=&\frac{9 (2\pi)^4 t^3}{2 N\NN^2 y^4}
	\left\{\begin{array}{ll} 2.08\cdot 10^{-3} & \mbox{for $\nu=2$}\\
	5.24\cdot 10^{-5} & \mbox{for $\nu=3$.} \end{array} \right.
\eea
In the limit $x\gg 1$ it decays like
\be
\left\lan |f_\pi|^2\right\ran(y\gg 1,t) \sim y^{-3-2\nu} t^3.
\ee

\subsection*{The unequal-time source functions}

We are not going to list all scalar unequal time correlators (UTC),
since there are no special problems involved in their
calculation, and they are not very illuminating anyway.
As an example, we present the UTC for $f_v$:
\be
\left\lan f_v(\bk,t)f_v^*(\bk,t')\right\ran
= \frac{(2\pi)^4 t r^2 }{N\NN^2 y^4} \int \!dx d\mu
	x^2 \left\{\left[y^2-x y \mu\right]^2 (\chi\vph)(\tilde{\chi}\tilde{\vph}) 
	+ \left[y^2-x y \mu\right]
	\left[x y \mu\right](\chi\vph)(\tilde{\vph}\tilde{\chi})\right\} ,
\ee
where we have additionally introduced $(\tilde{g}\tilde{h})\equiv g(xr)h(|\by-\bx|r)$
with $r\equiv t'/t$, while retaining the notation from the previous section, i.e.\ $x\equiv q t$ and $y\equiv k t$,
and $g(x) h(|\by-\bx|)\equiv (gh)$.

The correlators decay as power laws for large $r$. If we
parameterise them like $\lan f_i f_i^*\ran (y,r)\propto r^{-\ga_i}$,
we find for $r\gg1$
\be
\ga_\rho=3/2, \; \ga_p=3/2, \; \ga_v=3/2, \; \ga_\pi=5/2 .
\ee

The UTCs for the {scalar} seed variables $\Phi_{s}$ and $\Psi_{s}$
can then be pieced together using the above scalar variables as well
as the equations (\ref{e:Phis}) and (\ref{e:Psis}). For the
{vector} sources, we need to calculate the function $W(z,r)$.
We can use the fact that $W$ depends only on the magnitude
of $\bk$, but not on its direction: we choose the special direction
$\bk=(0,0,k)$. In that case we can for example use
$w_1^{(v)}$ in the expression
\be
W(kt,kt') = \frac{\left\lan w_1^{(v)}(k,t) w_1^{(v)*} (k,t')\right\ran}{k^2 \sqrt{tt'}} \, ,
\ee
based on Eqs.\ (\ref{e:Sis}) and (\ref{e:W}). From Eq.\ (\ref{eq:Tj0}) we can see that $w_i^{(v)}$ is given by
\be
w_i^{(v)} = T_{i0}^{(V)}=T_{i0}-\frac{k^i k^j}{k^2} T_{j0} \,.
\ee
Since for our choice of coordinates where $k_1=0$ we have that $w_1^{(v)}=T_{10}$,
the required correlator is then obtained as
\bea
k^2\sqrt{ts} W(kt,ks) &=& \left\lan T_{01} T_{01}^*\right\ran (k,t,s) =
A^2 t^2 s^2\int d^3\!q d^3\!p\, q_1 (-p_1) \vph(qt) \chi(|\bk-\bq|t)
\vph(ps)\chi(|-\bk-\bp|s) \left\lan\b^4_\mr{in}\right\ran \nonumber \\
&=& \frac{(2\pi)^3}{N\NN^2} t^2 s^2 \int d^3\!q\, q_1^2 \left(\chi(qs)\vph(|\bk-\bq|s)
-\vph(qs)\chi(|\bk-\bq|s)\right)\vph(qt)\chi(|\bk-\bq|t)\nonumber \\
&=& \frac{(2\pi)^3}{N\NN^2} \pi \frac{r^2}{t} \int dx\,d\mu\,x^4
(1-\mu^2)\left((\tilde{\chi}\tilde{\vph})-(\tilde{\vph}\tilde{\chi})\right) (\vph\chi) .
\eea
To perform the integration numerically, it can be advantageous to
change to an integration variable which is symmetric in $t$ and $s$,
\eg, $x\equiv qt \ra q\sqrt{ts}$.

The {tensor} type two-point functions are determined by $\tau^{(\pi)}_{ij}$.
We can use the same simplification as above, and for $\bk=(0,0,k)\equiv\bk_z$
we find
\bea
\lan \tau^{(\pi)}_{12}(\bk_z,t)\tau^{(\pi)*}_{12}(\bk_z,s) \ran &\equiv& T/\sqrt{ts} \quad \mbox{and} \\
\tau^{(\pi)}_{12}(\bk_z,t) &=& T_{12}(\bk_z,t) .
\eea
With the same variables as above, $T$ is then given by
\be
T(z,r) = \frac{(2\pi)^3}{N\NN^2} \frac{\pi}{2} r^{7/2} 
\int dx\,d\mu\,x^6 (1-\mu^2)^2 (\chi\chi) (\tilde{\chi}\tilde{\chi}) .
\ee

As explained in the main text, the UTC for the sources have to be diagonalized and their eigenvectors are then to be used as source terms in the linear perturbation equations of a Boltzmann solver.\\

\end{widetext}

\bibliography{CLNrefs}

\begin{thebibliography}{10}

\bibitem{Larson:2010gs}
D.~Larson {\em et~al.},
\newblock Astrophys. J. Suppl. {\bf 192}, 16 (2011), [arXiv:1001.4635],
\newblock 10.1088/0067-0049/192/2/16.

\bibitem{Komatsu:2010fb}
WMAP Collaboration, E.~Komatsu {\em et~al.},
\newblock Astrophys. J. Suppl. {\bf 192}, 18 (2011), [arXiv:1001.4538],
\newblock 10.1088/0067-0049/192/2/18.

\bibitem{Dunkley:2010ge}
J.~Dunkley {\em et~al.},
\newblock Astrophys.J. {\bf 739}, 52 (2011), [1009.0866],
\newblock 10.1088/0004-637X/739/1/52.

\bibitem{Reichardt:2011yv}
C.~Reichardt {\em et~al.},
\newblock Astrophys.J. {\bf 755}, 70 (2012), [1111.0932],
\newblock 10.1088/0004-637X/755/1/70.

\bibitem{Ade:2013hta}
Planck Collaboration, P.~Ade {\em et~al.},
\newblock 1303.5072.

\bibitem{Planck:2013kta}
Planck collaboration, P.~Ade {\em et~al.},
\newblock 1303.5075.

\bibitem{Kibble:1976sj}
T.~Kibble,
\newblock J.Phys.A {\bf A9}, 1387 (1976),
\newblock 10.1088/0305-4470/9/8/029.

\bibitem{Kibble:1980mv}
T.~Kibble,
\newblock Phys.Rept. {\bf 67}, 183 (1980),
\newblock 10.1016/0370-1573(80)90091-5.

\bibitem{VilenkinAndShellard}
A.~Vilenkin and E.~P.~S. Shellard,
\newblock {\em Cosmic Strings and Other Topological Defects} (Cambridge
  Monographs on Mathematical Physics, {Cambridge, UK}, 1994).

\bibitem{HindmarshAndKibble}
M.~Hindmarsh and T.~Kibble,
\newblock Rept.Prog.Phys. {\bf 58}, 477 (1995), [hep-ph/9411342],
\newblock 10.1088/0034-4885/58/5/001.

\bibitem{Durrer:2001cg}
R.~Durrer, M.~Kunz and A.~Melchiorri,
\newblock Phys.Rept. {\bf 364}, 1 (2002), [astro-ph/0110348],
\newblock 10.1016/S0370-1573(02)00014-5.

\bibitem{Durrer:1994zza}
R.~Durrer,
\newblock Fund.Cosmic Phys. {\bf 15}, 209 (1994), [astro-ph/9311041].

\bibitem{Durrer2}
R.~Durrer, A.~Gangui and M.~Sakellariadou,
\newblock Phys.Rev.Lett. {\bf 76}, 579 (1996), [astro-ph/9507035],
\newblock 10.1103/PhysRevLett.76.579.

\bibitem{Durrer3}
R.~Durrer, M.~Kunz and A.~Melchiorri,
\newblock Phys.Rev. {\bf D59}, 123005 (1999), [astro-ph/9811174],
\newblock 10.1103/PhysRevD.59.123005.

\bibitem{Turok1}
U.-L. Pen, U.~Seljak and N.~Turok,
\newblock Phys.Rev.Lett. {\bf 79}, 1611 (1997), [astro-ph/9704165],
\newblock 10.1103/PhysRevLett.79.1611.

\bibitem{Turok2}
U.-L. Pen, D.~N. Spergel and N.~Turok,
\newblock Phys.Rev. {\bf D49}, 692 (1994),
\newblock 10.1103/PhysRevD.49.692.

\bibitem{Mark1}
M.~Hindmarsh, C.~Ringeval and T.~Suyama,
\newblock Phys.Rev. {\bf D80}, 083501 (2009), [0908.0432],
\newblock 10.1103/PhysRevD.80.083501.

\bibitem{Mark2}
M.~Hindmarsh, C.~Ringeval and T.~Suyama,
\newblock Phys.Rev. {\bf D81}, 063505 (2010), [0911.1241],
\newblock 10.1103/PhysRevD.81.063505.

\bibitem{Shellard}
D.~Regan and E.~Shellard,
\newblock Phys.Rev. {\bf D82}, 063527 (2010), [0911.2491],
\newblock 10.1103/PhysRevD.82.063527.

\bibitem{Dani}
D.~G. Figueroa, R.~R. Caldwell and M.~Kamionkowski,
\newblock Phys.Rev. {\bf D81}, 123504 (2010), [1003.0672],
\newblock 10.1103/PhysRevD.81.123504.

\bibitem{Hill:1986mn}
C.~T. Hill, D.~N. Schramm and T.~P. Walker,
\newblock Phys.Rev. {\bf D36}, 1007 (1987),
\newblock 10.1103/PhysRevD.36.1007.

\bibitem{KibbleCosmicRays}
A.~Gill and T.~Kibble,
\newblock Phys.Rev. {\bf D50}, 3660 (1994), [hep-ph/9403395],
\newblock 10.1103/PhysRevD.50.3660.

\bibitem{Dimopoulos:1997df}
K.~Dimopoulos,
\newblock Phys.Rev. {\bf D57}, 4629 (1998), [hep-ph/9706513],
\newblock 10.1103/PhysRevD.57.4629.

\bibitem{DufauxFigueroaBellido}
J.-F. Dufaux, D.~G. Figueroa and J.~Garcia-Bellido,
\newblock Phys.Rev. {\bf D82}, 083518 (2010), [1006.0217],
\newblock 10.1103/PhysRevD.82.083518.

\bibitem{JonesSmith:2007ne}
K.~Jones-Smith, L.~M. Krauss and H.~Mathur,
\newblock Phys.Rev.Lett. {\bf 100}, 131302 (2008), [0712.0778],
\newblock 10.1103/PhysRevLett.100.131302.

\bibitem{Fenu:2009qf}
E.~Fenu, D.~G. Figueroa, R.~Durrer and J.~Garcia-Bellido,
\newblock JCAP {\bf 0910}, 005 (2009), [0908.0425],
\newblock 10.1088/1475-7516/2009/10/005.

\bibitem{Giblin:2011yh}
J.~Giblin, John~T., L.~R. Price, X.~Siemens and B.~Vlcek,
\newblock JCAP {\bf 1211}, 006 (2012), [1111.4014],
\newblock 10.1088/1475-7516/2012/11/006.

\bibitem{Figueroa:2012kw}
D.~G. Figueroa, M.~Hindmarsh and J.~Urrestilla,
\newblock Phys.Rev.Lett. {\bf 110}, 101302 (2013), [1212.5458],
\newblock 10.1103/PhysRevLett.110.101302.

\bibitem{Vilenkin:1981bx}
A.~Vilenkin,
\newblock Phys.Lett. {\bf B107}, 47 (1981),
\newblock 10.1016/0370-2693(81)91144-8.

\bibitem{Vachaspati:1984gt}
T.~Vachaspati and A.~Vilenkin,
\newblock Phys.Rev. {\bf D31}, 3052 (1985),
\newblock 10.1103/PhysRevD.31.3052.

\bibitem{Olmez2010}
S.~Olmez, V.~Mandic and X.~Siemens,
\newblock Phys. Rev. {\bf D81}, 104028 (2010).

\bibitem{Blanco-Pillado:2013qja}
J.~J. Blanco-Pillado, K.~D. Olum and B.~Shlaer,
\newblock 1309.6637.

\bibitem{Urrestilla:2011gr}
J.~Urrestilla, N.~Bevis, M.~Hindmarsh and M.~Kunz,
\newblock JCAP {\bf 1112}, 021 (2011), [1108.2730],
\newblock 10.1088/1475-7516/2011/12/021.

\bibitem{Binetruy:2012ze}
P.~Binetruy, A.~Bohe, C.~Caprini and J.-F. Dufaux,
\newblock JCAP {\bf 1206}, 027 (2012), [1201.0983],
\newblock 10.1088/1475-7516/2012/06/027.

\bibitem{Sanidas:2012ee}
S.~Sanidas, R.~Battye and B.~Stappers,
\newblock Phys.Rev. {\bf D85}, 122003 (2012), [1201.2419],
\newblock 10.1103/PhysRevD.85.122003.

\bibitem{Bevis:2004wk}
N.~Bevis, M.~Hindmarsh and M.~Kunz,
\newblock Phys.Rev. {\bf D70}, 043508 (2004), [astro-ph/0403029],
\newblock 10.1103/PhysRevD.70.043508.

\bibitem{Urrestilla:2007sf}
J.~Urrestilla, N.~Bevis, M.~Hindmarsh, M.~Kunz and A.~R. Liddle,
\newblock JCAP {\bf 0807}, 010 (2008), [0711.1842],
\newblock 10.1088/1475-7516/2008/07/010.

\bibitem{Ade:2013xla}
Planck Collaboration, P.~Ade {\em et~al.},
\newblock [1303.5085].

\bibitem{Durrer:2001xu}
R.~Durrer, M.~Kunz and A.~Melchiorri,
\newblock Phys.Rev. {\bf D63}, 081301 (2001), [astro-ph/0010633],
\newblock 10.1103/PhysRevD.63.081301.

\bibitem{Turok:1991qq}
N.~Turok and D.~N. Spergel,
\newblock Phys.Rev.Lett. {\bf 66}, 3093 (1991),
\newblock 10.1103/PhysRevLett.66.3093.

\bibitem{GarciaBellido:2010if}
J.~Garcia-Bellido, R.~Durrer, E.~Fenu, D.~G. Figueroa and M.~Kunz,
\newblock Phys.Lett. {\bf B695}, 26 (2011), [1003.0299],
\newblock 10.1016/j.physletb.2010.11.031.

\bibitem{Felder99}
G.~N. Felder {\em et~al.},
\newblock Phys.Rev.Lett. {\bf 87}, 011601 (2001), [hep-ph/0012142],
\newblock 10.1103/PhysRevLett.87.011601.

\bibitem{Kunz:1996ka}
M.~Kunz and R.~Durrer,
\newblock Phys.Rev. {\bf D55}, 4516 (1997), [astro-ph/9612202],
\newblock 10.1103/PhysRevD.55.R4516.

\bibitem{mkthesis}
M.~Kunz,
\newblock Th\`ese de doctorat: Univ. Gen\`eve  (1999).

\bibitem{Jaffe:1993tt}
A.~H. Jaffe,
\newblock Phys.Rev. {\bf D49}, 3893 (1994), [astro-ph/9311023],
\newblock 10.1103/PhysRevD.49.3893.

\bibitem{Durrer:1997ep}
R.~Durrer and M.~Kunz,
\newblock Phys.Rev. {\bf D57}, R3199 (1998), [astro-ph/9711133],
\newblock 10.1103/PhysRevD.57.3199.

\bibitem{Durrer:1998rw}
R.~Durrer, M.~Kunz and A.~Melchiorri,
\newblock Phys.Rev. {\bf D59}, 123005 (1999), [astro-ph/9811174],
\newblock 10.1103/PhysRevD.59.123005.

\bibitem{Doran:2003sy}
M.~Doran,
\newblock JCAP {\bf 0510}, 011 (2005), [astro-ph/0302138],
\newblock 10.1088/1475-7516/2005/10/011.

\bibitem{Spergel:1997vq}
D.~N. Spergel and M.~Zaldarriaga,
\newblock Phys.Rev.Lett. {\bf 79}, 2180 (1997), [astro-ph/9705182],
\newblock 10.1103/PhysRevLett.79.2180.

\bibitem{Scodeller:2009iu}
S.~Scodeller, M.~Kunz and R.~Durrer,
\newblock Phys.Rev. {\bf D79}, 083515 (2009), [0901.1845],
\newblock 10.1103/PhysRevD.79.083515.

\bibitem{Durrer:2008aa}
R.~Durrer,
\newblock {\em The Cosmic Microwave Background} (Cambridge University Press,
  {Cambridge, UK}, 2008).

\bibitem{wPlanck}
Planck legacy archive,
\newblock http://www.sciops.esa.int/index.php... \\
  ...?project=planck\&page=Planck\underline{~}Legacy\underline{~}Archive.

\bibitem{wCore}
Core proposal,
\newblock White paper arXiv:1102.2181, http:www.core-mission.org.

\bibitem{wPrism}
Prism proposal,
\newblock White papers arXiv:1306.2259, arXiv:1310.1554,
  http:www.prism-mission.org.

\end{thebibliography}

\bibliographystyle{h-physrev4}

\end{document}